\documentclass[useAMS,usenatbib]{mn2e}

\usepackage{graphicx}
\usepackage{times}%
\usepackage{natbib}

\renewcommand{\vec}[1]{ {\bmath #1} } 

\def\ltsima{$\; \buildrel < \over \sim \;$}
\def\simlt{\lower.5ex\hbox{\ltsima}}
\def\gtsima{$\; \buildrel > \over \sim \;$}
\def\simgt{\lower.5ex\hbox{\gtsima}}

\title[Dark matter structures in early dark energy cosmologies]
{The impact of Early Dark Energy on non-linear structure formation}
\author[Grossi \& Springel]
{Margherita Grossi$^{1}$, Volker Springel$^{1}$\\
$^1$ Max-Planck Institut fuer Astrophysik,
Karl-Schwarzschild Strasse 1, D-85748 Garching, Germany\\
(margot@mpa-garching.mpg.de),(volker@mpa-garching.mpg.de)\\
}

\begin{document}

\date{Accepted ???. Received ???; in original form }

\pagerange{\pageref{firstpage}--\pageref{lastpage}} \pubyear{2007}

\maketitle

\label{firstpage}

\begin{abstract}
  We study non-linear structure formation in high-resolution simulations of
  Early Dark Energy (EDE) cosmologies and compare their evolution with the
  standard $\Lambda$CDM model. In Early Dark Energy models, the impact on
  structure formation is expected to be particularly strong because of the
  presence of a non-negligible dark energy component even at very high
  redshift, unlike in standard models that behave like matter-dominated
  universes at early times.  In fact, extensions of the spherical top-hat
  collapse model predict that the virial overdensity and linear threshold
  density for collapse should be modified in EDE model, yielding significant
  modifications in the expected halo mass function. Here we present numerical
  simulations that directly test these expectations.  Interestingly, we find
  that the Sheth \& Tormen formalism for estimating the abundance of dark
  matter halos continues to work very well in its standard form for the Early
  Dark Energy cosmologies, contrary to analytic predictions. The residuals are
  even slightly smaller than for $\Lambda$CDM. We also study the virial
  relationship between mass and dark matter velocity dispersion in different
  dark energy cosmologies, finding excellent agreement with the normalization
  for $\Lambda$CDM as calibrated by \citet{Evrard2008}.  The earlier growth of
  structure in EDE models relative to $\Lambda$CDM produces large differences
  in the mass functions at high redshift. This could be measured directly by
  counting groups as a function of the line-of-sight velocity dispersion,
  skirting the ambiguous problem of assigning a mass to the halo. Using dark
  matter substructures as a proxy for member galaxies, we demonstrate that
  even with 3-5 members sufficiently accurate measurements of the halo
  velocity dispersion function are possible. Finally, we determine the
  concentration-mass relationship for our EDE cosmologies. Consistent with the
  earlier formation time, the EDE halos show higher concentrations at a given
  halo mass.  We find that the magnitude of the difference in concentration is
  well described by the prescription of \citet{ENS2001} for estimating halo
  concentrations.

\end{abstract}

\begin{keywords}
early universe --  cosmology: theory -- galaxies: formation
\end{keywords}


\section{Introduction} \label{sect:intro}

Arguably the most surprising result of modern cosmology is that all matter
(including both atoms and non-baryonic dark matter) accounts for only a
quarter of the total energy density of the Universe today, while the rest is
contributed by a {\em dark energy} field.  In 1999, observations of type Ia
supernovae by the Supernovae Cosmology Project \citep{Riess1999,Riess2004} and
the relative accurate measurements of the distances to this objects
\citep{Perlmutter1999,Kowalski2008} demonstrated that the expansion of the
Universe is accelerated today; there hence exists a mysterious force that acts
against the pull of gravity. Nowadays, the inference that this is caused by
dark energy can be made with significant confidence, as the observational
evidence has further firmed up.  In fact, we have good reason to believe that
we live in a flat universe with an upper limit of $\Omega_m \leq 0.3$ for the
matter density today, based on cosmic microwave background measurements and a
host of other observational probes \citep[e.g.]{WMAP5}.  These observations
yield a consistent picture, the so-called concordance cosmology, and are in
agreement with predictions of the inflationary theory.

The physical origin of dark energy is however unknown and a major puzzle for
theoretical physics. A nagging outstanding problem is that most quantum field
theories predict a huge cosmological constant from the energy of the quantum
vacuum, up to 120 orders of magnitude too large. There is hence no simple
natural explanation for dark energy, and one has to be content with
phenomenological models at this point. Two proposed forms of dark energy are
the cosmological constant, a constant energy density filling space
homogeneously, and scalar fields such as quintessence. In particular,
`tracking quintessence' models attempt to alleviate the coincidence problem of
the cosmological constant model. More exotic models where the dark energy
couples to matter fields or can cluster itself have also been proposed.

In light of the many theoretical possibilities, the hope is that future
observational constraints on dark energy will enable progress in the
understanding of this puzzling phenomenon. This requires the exploitation of
the subtle influence of dark energy on structure formation, both on linear and
non-linear scales. As the expected effects are generally small for many of the
viable dark energy scenarios, it is crucial to be able to calculate structure
formation in dark energy cosmologies with sufficient precision to tell the
different models apart, and to be able to correctly interpret observational
data. For example, in order to use the abundance of clusters of galaxies at
different epochs to measure the expansion history of the universe, one needs
to reliably know how the cluster mass function evolves with time in different
dark energy cosmologies. Numerical simulations are the most accurate tool
available to obtain the needed theoretical predictions, and they are also
crucial for testing the results of more simplified analytic calculations.

In this study, we carry out such non-linear simulations for a particular class
of dark energy cosmologies, so-called Early Dark Energy (EDE) models where
dark energy might constitute an observable fraction of the total energy
density of our Universe at the time of matter radiation equality or even
big-bang nucleosynthesis. While in the cosmological constant scenario, the
fraction in dark energy is negligible at high redshift, in such models the
energy fraction is a few per cent during recombination and structure
formation, which introduces interesting effects due to dark energy already at
high redshift. In particular, for an equal amplitude of clustering today, we
expect structures to form earlier in such cosmologies than in
$\Lambda$CDM. This could be useful to alleviate the tension between a low
$\sigma_8$ normalization suggested by current observational constraints from
the CMB on one hand, and the observations of relatively early reionization and
the existence of a population of massive halos present already at high
redshift on the other hand.

Recently, \citet{Bartelmann2006} studied two particular EDE models, evaluating
the primary quantities relevant for structure formation, such as the linear
growth factor of density perturbation, the critical density for spherical
collapse and the overdensity at virialization, and finally the halo mass
function.  In the two models analyzed, they found that the effect of EDE on
the geometry of the Universe is only moderate, for example, distance measures
can be reduced by $~8\%$. Assuming the same expansion rate today, such models
are younger compared to $\Lambda$CDM. At early times, the age of the universe
should differ by approximately $5-10 \% $.

However, when \citet{Bartelmann2006} repeated the calculation of the spherical
collapse model in the EDE cosmology, a few nontrivial modifications appeared.
The evolution of a homogeneous, spherical overdensity can be traced utilizing
both the virial theorem and the energy conservation between the collapse and
the turn around time \citep[see
also][]{Lacey1993,Wang1998}. \citet{Bartelmann2006} obtained the value of the
virial overdensity as a function of the collapse redshift, translating the
effect of the early dark energy in an extra contribution to the potential
energy at early times.  They found that the virial overdensity should be
slightly enlarged by EDE, because a faster expansion of the universe means
that, by the time a perturbation has turned around and collapsed to its final
radius, a larger density contrast has been produced.  However, at the same
time they found that the linearly extrapolated density contrast corresponding
to the collapsed object should be significantly reduced.

These two results based on analytic expectations have a pronounced influence
on the predicted mass function of dark matter halos.  In EDE models, the
cluster population expected from the Press-Schechter or Sheth-Tormen formalism
grows considerably relative to $\Lambda$CDM, as a result of the lowered value
of the critical linear density contrast $\delta_c$ for collapse. This effect
can be compensated for by lowering the normalization parameter $\sigma_8$ in
order to obtain the same abundance of clusters today. In this case, one would
however still expect a higher cluster abundance in EDE at high redshift, due
to the earlier growth of structure in this model.

An open question is whether the EDE really participates in the virialization
process in the way assumed in the analytic modeling.  Similarly, it is not
clear whether the excursion set formalism of Sheth \& Tormen yields an equally
accurate description of the non-linear mass function of halos in EDE
cosmologies as in $\Lambda$CDM.  Because accurate theoretical predictions for
the halo mass function are a critical ingredient for constraining cosmological
parameters (in particular $ \Omega_m$ and $ \Omega_{\Lambda}$) as well as
models of galaxy formation, it is important to test these predictions for the
EDE cosmology in detail with numerical N-body simulations. In particular, we
want to probe whether the fraction $f$ of matter ending up in objects larger
than a given mass $M$ at some redshift $z$ can be found by {\em only} looking
at the properties of the linearly evolved density field at this epoch, using
the ordinary ST formalism, or whether there is some dependence on redshift,
power spectrum or dark energy parameters, as suggested by
\citet{Bartelmann2006}.

A further interest in EDE cosmologies stems from the fact that for a given
$\sigma_8$, the EDE models predict a substantially slower evolution of the
halo population than in the $ \Lambda$CDM model. This could explain the higher
normalization cosmology expected from cluster studies relative to analysis of
the CMB. The value of $\sigma_8$, for a given cosmology, provides also a
measure of the expected biasing parameter that relates the galaxy and the mass
distribution. The early dark energy cosmologies could hence reduce the current
mild tension between cluster data and the CMB observations. We note that halos
in cosmologies with EDE are also expected to be more concentrated than in
$\Lambda$CDM; because the density of the Universe was greater at early times, objects
that virialized at high redshift are more compact than those that virialized
more recently.

Previous numerical simulations of a quintessence component with a changing
equation of state (EOS) explored two particular potentials: SUGRA and Ratra
Peebles (RP), which differ because RP has a more smoothly decreasing $w$ and
consequently a very different evolution in the past. Both \cite{Linder2003}
and \cite{Klypin2003} analyzed the influence of the dark energy on the halo
mass function in order to extrapolate the abundance of structure at different
epochs and to compare it with existent theoretical models. They used different
numerical codes: the publicly available code {\small GADGET}, in the first
project, and the Adaptive Refinement Tree code \citep{ARTcode}, in the
second. They concluded that the best way to understand which dark energy Universe fit
the observations best is to look at the growth history of halos and
the evolution of their properties with time. \citet{Dolag2004}
focused on the modification of the concentration parameter with mass and
redshift, for the same cosmologies, based on high resolution simulations of a
sample of massive halos. A limited number of numerical studies also considered
the possibility of a coupling of the dark energy field with dark matter
\citep{Mainini2003,Maccio}.

In this paper, we carry out several high resolution simulations of dark energy
cosmologies in order to accurately measure the quantitative impact of early
dark energy on abundance and structure of dark matter halos. To this end, we
in particular measure halo mass functions and evaluate the
agreement/disagreement with different analytic fitting functions. We also test
how well the growth of the mass function can be tracked with dynamical measure
based on the velocity dispersion of dark matter substructures, which can serve
as a proxy for the directly measurable line-of-sight motion of galaxies or
line widths in observations, and gets around the usual ambiguities arising
from different possible mass definitions for halos. Finally, we also present
measurements of halo concentrations, and of the relation between dark matter
velocity dispersion and halo mass. While finalizing this paper,
\citet{Francis2008} submitted a preprint which also studies numerical
simulations of EDE cosmologies. Their work provides a different analysis and
is complementary to our study, but it reaches similar basic conclusions about
the halo mass function.

This paper is organized as follows. After a brief introduction to the Early
Dark Energy models in Section 2, we present the simulations and also give
details on our numerical methods in Section~3. In Section~4, we study the mass
function of halos for the different cosmologies, and as a function of
redshift. Then, in Section~5 we investigate the properties of halos by
studying the virial relation between mass and dark matter velocity dispersion,
as well as the mass--concentration relationship. In Section~6, we consider the
velocity distribution function and prospects for measuring it in observations.
Finally, we discuss our results and present our conclusions in Section 7.

\section{Early dark energy models}

The influence of dark energy on the evolution of the Universe is
governed by its 
equation of state,
\begin{equation}
	p = w \rho c^2.
\end{equation}
A cosmological constant has $w_{\Lambda}=-1$ at all redshift, while a
distinctive feature of the Early Dark Energy (hereafter EDE) models as well as
of other models such as quintessence is that their equation of state
parameter, $w_{\rm de}(z)$, varies during cosmic history.

Negative pressure at all times implies that the energy density parameter will
fall to zero very steeply for increasing redshift. If, however, we allow the
equation of state parameter to rise above zero, we can construct models in
which $ \Omega_{\rm de}(z)$ has a small positive value at all epochs,
depending on the cosmological background model we adopt. While canonical dark
energy models with near constant behaviour for $w$ do not predict any
substantial dark energy effect at $z>2$, in such EDE models the contribution
of dark energy to the cosmic density can be of order of a few percent even at
very high redshift.

We are here investigating this interesting class of models which are
characterized by a low but non-vanishing dark energy density at early
times. Note that while the acceleration of the expansion of the Universe is a
quite recent phenomenon, the dark energy responsible for this process could
have an old origin. In fact, field theoretical models have been constructed
that generically cause such a dynamical behaviour
\citep{RatraPeebles1988,Wetterich1988,Ferreira1998,Liddle1999}.

\citet{Wetterich2004} proposed a useful parameterization of a family of
cosmological models with EDE in terms of three parameters:
\begin{itemize}
\item the amount of dark energy today, $ \Omega_{\rm de,0}$ (we assume a flat
  universe, so $ \Omega_{m,0} = 1-\Omega_{\rm de,0}$),
\item the equation-of-state parameter $w_0$ today, and
\item an average value $ \Omega_{\rm de,e}$ of the energy density parameter at
  early times (to which it asymptotes for $z \mapsto \infty$).
\end{itemize}

\begin{figure}
\begin{center}
\includegraphics[width = 0.4\textwidth]{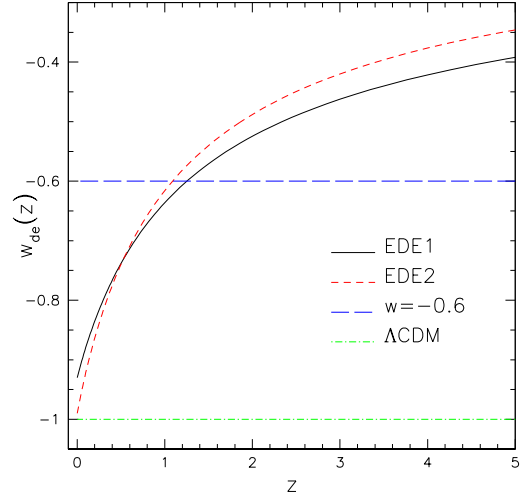}
\caption{Equation of state parameter $w$ shown as a function of redshift for
  the four different cosmological models considered in this work. In the two
  early dark energy models EDE1 and EDE2, shown with black solid and red
  dotted lines respectively, the value of $w$ today is close to that of
  $\Lambda$CDM, but the amount of dark energy at early times is non
  vanishing, as described by the parameterization (\ref{eq:www}).  }
\label{fig:wz}
\end{center}
\end{figure}

Figure \ref{fig:wz} shows the redshift evolution of the equation-of-state
parameter in the four different cosmologies that we examine in this study. As
can be noticed, the EDE models approach the cosmological constant scenario at
very low redshift.  We can compute the equation-of-state parameter for these
early dark energy models from the fitting formula:
\begin{equation}
	w(z)=\frac{w_0}{\left(1+by\right)^{2}},
	\label{eq:www}
\end{equation}
where
\begin{equation}
	b=-\frac{3w_0}{\ln\left(\frac{1/\Omega_{\rm de,e}}{\Omega_{\rm
              de,e}}\right)+\ln\left(\frac{1-\Omega_{m,0}}{\Omega_{m,0}}\right)} \label{eq:bpar},
\end{equation}
and $y=\ln\left(1+z\right)=-\ln a$. The parameter $b$ characterizes the time
at which an approximately constant equation-of-state changes its behaviour.
 
\begin{figure}
\begin{center}
\includegraphics[width = 0.4\textwidth]{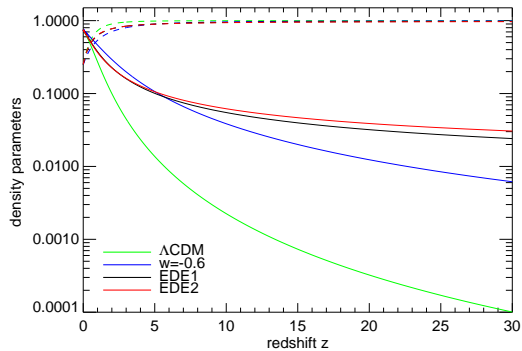}
\caption{Evolution of the density parameters $\Omega_{m}(z)$ (dashed lines)
  and $\Omega_{\rm de}(z)$ (solid line) for the four cosmological models
  studied in this work. At redshift $z=30$, the dark energy contribution is
  orders of magnitude higher for EDE models compared with a
  $\Lambda$CDM cosmology.  }
\label{fig:densitypar}
\end{center}
\end{figure}

In Figure \ref{fig:densitypar}, we plot the evolution of the matter and energy
density parameters up to redshift $z=30$. The dark energy parameter for EDE
models evolves relatively slowly with respect to a standard $\Lambda$CDM
cosmology. In fact, the critical feature of this parameterization is a
non-vanishing dark energy contribution during recombination and structure
formation \citep[see also][]{Doran2001}:
\begin{equation}
	\bar{\Omega}_{\rm de,sf} = -\ln\, a_{\rm eq}^{-1}\int^{0}_{\ln
          a_{\rm eq}} \Omega_{\rm de}\left(a\right){\rm d} \ln a .
\label{eq:Omegasf}
\end{equation}
For sufficiently low $\Omega_{\rm de,e}$, the EDE models reproduce quite well
the accelerated cosmic expansion in the present-day Universe and they can be
fine-tuned to agree both with low-redshift observations and CMB temperature
fluctuation results \citep{Doran2005,Doran2007}.

\section{Numerical simulations}

\begin{table}
\begin{center}
\begin{tabular}{l|rrrrrr}
\hline
         & $ \Omega_{m,0}$ & $ \Omega_{{\rm de},0}$ & $h_0$ & $ \sigma_{8}$ &
         $w_{0}$ & $ \Omega_{{\rm de},e}$ \\
\hline
$\Lambda$CDM & 0.25 & 0.75 &  0.7 & 0.8 & -1.   & 0.         \\
DECDM        & 0.25 & 0.75 &  0.7 & 0.8 & -0.6  & 0.         \\
EDE1         & 0.25 & 0.75 &  0.7 & 0.8 & -0.93 & $2\times 10^{-4}$ \\
EDE2         & 0.25 & 0.75 &  0.7 & 0.8 & -0.99 & $8\times 10^{-4}$ \\
\hline
\end{tabular}
\end{center}
\caption{Parameters of the N-Body simulations. The parameter 
  $\Omega_{\rm de,e}$ describes the amount of dark energy at early times,
  see equation (\ref{eq:bpar}). This value, together with
  $w_0$, the value of the equation state parameter today, and 
  $\Omega_{\rm de,0}$, the amount of dark energy today, completely describes
  our EDE models. }
\label{tab:param}
\end{table}

We performed a series of cosmological N-body simulations for two early dark
energy models `EDE1' and `EDE2', which have $w_0=-0.93$ and $w_0=-0.99$,
respectively, and a dark energy density at early times of about $10^{-4}$ (see
Tab.~\ref{tab:param}). For comparison, we have also calculated a model `DECDM'
with constant equation of state parameter equal to $w=-0.6$, and a
conventional $\Lambda$CDM reference model. We shall refer with these labels to
the different models throughout the paper.

In all our models, the matter density parameter today was chosen as
$\Omega_m=0.25$, and we consider a flat universe. The Hubble parameter is
$h=H_0 / (100\,{\rm km\, s^{-1}Mpc^{-1}})= 0.7$ and we assume Gaussian density
fluctuations with a scale-invariant primordial power spectrum.  The
normalization of the linear power spectrum extrapolated to $z=0$ is
$\sigma_8=0.8$ for all our simulations in order to match the observed
abundance of galaxy clusters today, irrespective of the cosmology. We also
used the same spectral index $n=-1$ throughout in order to focus our attention
on possible differences due to the dark energy contribution alone. For these
choices, the models EDE1 and EDE2 are almost degenerate, but their proximity
serves as a useful test for how well differences in the results can be
detected even for small variations in the EDE parameters. This gives a useful
illustration on how well one can hope to be able to distinguish them in
practice and provides realistic data for testing the discriminative power of
specific statistics.

For our largest calculations we used $512^3$ particles in boxes of volume
$100^3\,h^{-3}{\rm Mpc}^3$, resulting in a mass resolution of $m_p=5.17 \times
10^{8}~h^{-1}M_{\odot}$ and a gravitational softening length of
$\epsilon=4.2\,h^{-1}{\rm kpc}$, kept fixed in comoving coordinates.  All the
simulations were started at redshift $z_{\rm init}=49$, and evolved to the
present.  For the simulations, we adapted the cosmological code {\small
  GADGET-3} \citep[based on][]{SpringelGadget1,SpringelGadget2} and the
initial condition code {\small N-GENIC}, in order to allow simulations with a
time-variable equation of state.  These simulations can be used to determine
the mass function also in the high-mass tail with reasonably small cosmic
variance error, while at the same time probing down to interestingly small
mass scales.

\begin{figure}
\begin{center}
\includegraphics[width = 0.4\textwidth]{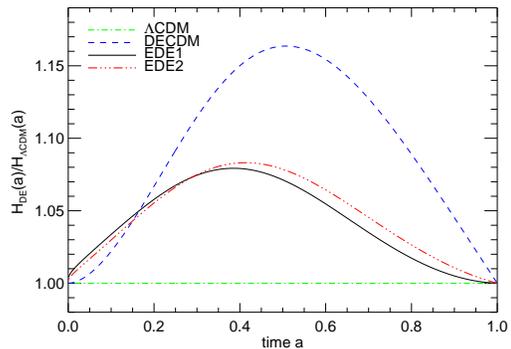}
\caption{Hubble expansion rate for the models studied in this work. All models
  are normalized with respect to the reference $\Lambda$CDM case. In the
  models EDE1, EDE2, and in the model with constant $w$, the expansion rate of
  the universe is higher at early times. This has a strong effect on the
  evolution of the growth factor.  }
\label{fig:expansion}
\end{center}
\end{figure}

In Figure \ref{fig:expansion}, we plot the expansion function of the EDE
models relative to the $\Lambda$CDM case. We note that the only modification
required in the simulation code was to update the expression for calculating
the Hubble expansion rate, which needs to include the quintessence
component. This term enters in both the kinematics and the dynamics of the
cosmological models.

According to the Friedmann equation within a flat universe we have
\begin{equation}
	H(a) = H_0 \left[\frac{\Omega_{m,0}}{a^3}+\Omega_{\rm de,0}\, \exp
          \left( -3 \int  \left[1+w \left(a \right)
            \right]{\rm d} \ln a\right) \right]^{1/2}.
\end{equation}
The density of dark energy changes with the scale factor as:
 \begin{equation}
\Omega_{\rm de}(z) =  \Omega_{\rm de,0} \exp\left( -3 \int d \ln
   a\left[1+w\left(a\right)\right]\right),\label{eq:DEterm}
 \end{equation}
instead of simply being equal to $\Omega_{\rm de,0}$, as in the usual
scenario. For $w=-1$, the behaviour of a cosmological constant is recovered.

If we interpret the modified expansion rate as being due to $w(z)$, as
defined in equation~(\ref{eq:www}), we find:
\begin{equation}
	H^2\left(z\right)/H_{0}^2=\Omega_{\rm
          de,0}\left(1+z\right)^{3+3\bar{w}_{h}\left(z\right)}+\Omega_{m,0}\left(1+z\right)^{3},
\end{equation}
where
\begin{equation}
	\bar{w}_{h}\left(z\right)=\frac{w_{0}}{1+b \ln
          \left(1+z\right)},
\end{equation}
and $b$ is given by the Eqn.~(\ref{eq:bpar}).

We can see that effectively the EDE models predict the observed effect of an
acceleration in the expansion rate, and this has consequences on the global
geometry of the Universe. We note that the dark energy term in
Eqn.~(\ref{eq:DEterm}) just parametrizes our ignorance concerning the physical
mechanism leading to an increase in expansion rate. However, once the
dependence of $H$ on the scale factor is fixed, the mathematical problem of
calculating structure growth is then unambiguously defined.

\begin{figure}
\begin{center}
\includegraphics[width = 0.4\textwidth]{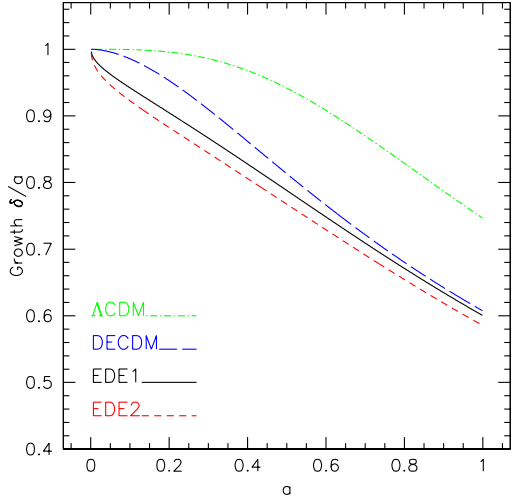}
\caption{Ratio of the growth factor of linear density perturbations and the
  scale factor $a$, as a function of $a$. The four models are described in
  Table~\ref{tab:param}. The curves are normalized to unity at early times,
  i.e.~we here assume that the starting density contrast is the same in the
  four cosmologies. The models EDE1 and EDE2 show a significant difference in
  the growth factor evolution even with small energy density at high redshift:
  structures have to grow earlier to reach the same abundance as the
  $\Lambda$CDM model today.}
\label{fig:grow}
\end{center}
\end{figure}

The evolution of $\Omega_{\rm de,a}$ affects not only the expansion rate of
the background but also the formation of structures.  The primary influence of
dark energy on the growth of matter density perturbations is however indirect
and arises through the sensitive dependence of structure growth on the
expansion rate of the universe. In Figure \ref{fig:grow}, we show the linear
growth factor $D$ divided by the scale factor $ D /a $ as a function of time
for all our models. All curves are normalized so that they start from unity at
early times.

In order to rescale the power spectrum of matter fluctuations to the redshift
of the initial conditions ($z=49$ for all simulations), we introduced in our
initial condition code the calculation of the growth factor for a general
equation-of-state as given by \citet{Linder2003}:
\begin{equation}
	D''+\frac{3}{2}\left[1-\frac{w\left(a\right)}{1+X\left(a\right)}\frac{D'}{a}+\frac{3}{2}\frac{X\left(a\right)}{1+X\left(a\right)}\frac{D}{a^2}\right]
        =0,
\end{equation}
where $X\left(a\right)$ is the ratio of the matter density to the energy density:
\begin{equation}
	X\left(a\right)=\frac{\Omega_{m,0}}{\Omega_{\rm de,0}} \exp
        \left[-3 \int^{1}_{a} d \ln a' w\left(a'\right) \right],
\end{equation}
and we allowed for a time-dependent equation of state, $w(a)$.  Here we define
the growth factor as the ratio $D =
\delta\left(a\right)/\delta\left(a_i\right)$ of the perturbation amplitude at
scale factor $a$ relative to the one at $a_i$, and we use the normalization
condition $D(a_{\rm eq})=a_{\rm eq}$.

\begin{figure}[H]
\centering
\includegraphics[width=0.45\textwidth]{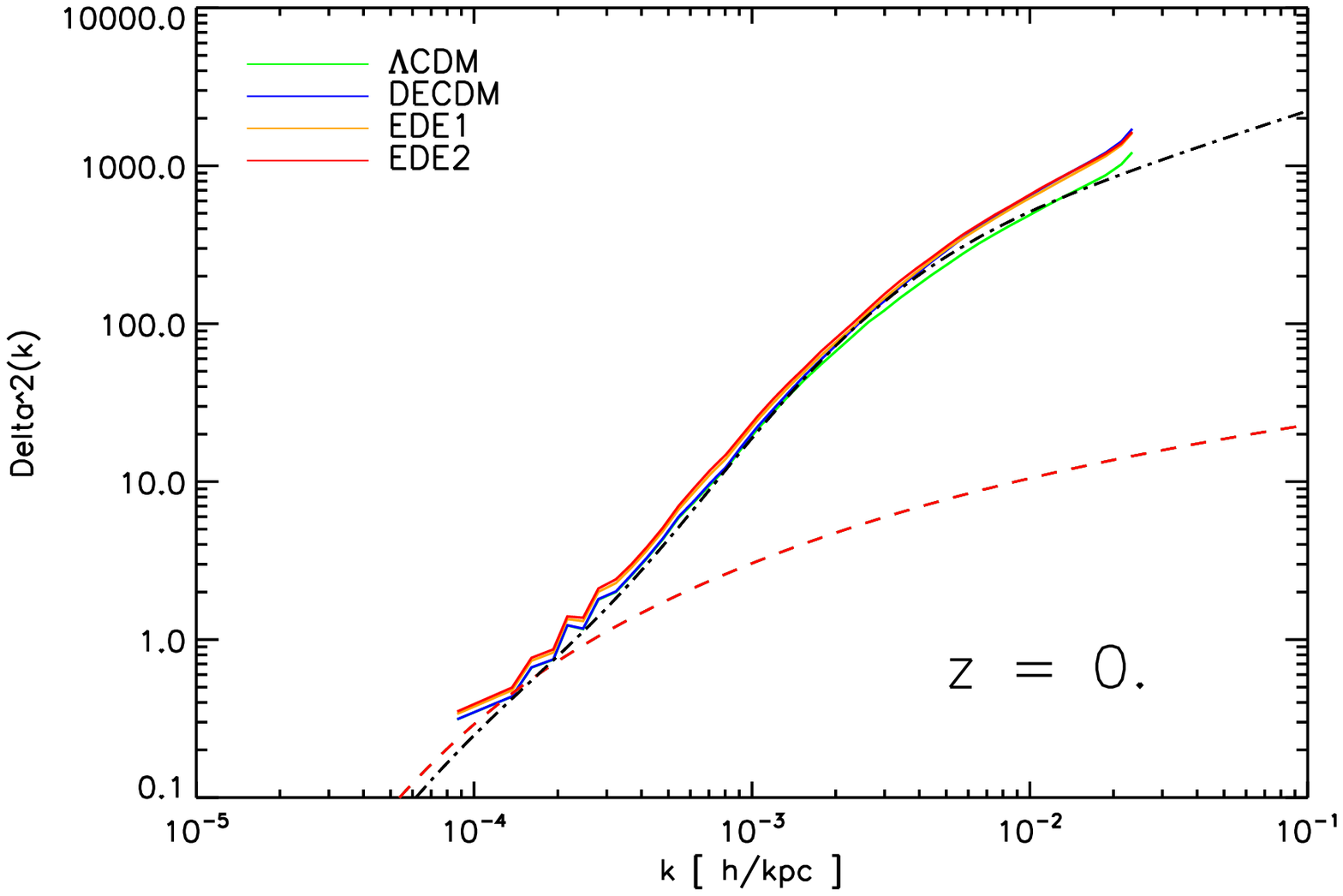}\\
\includegraphics[width=0.45\textwidth]{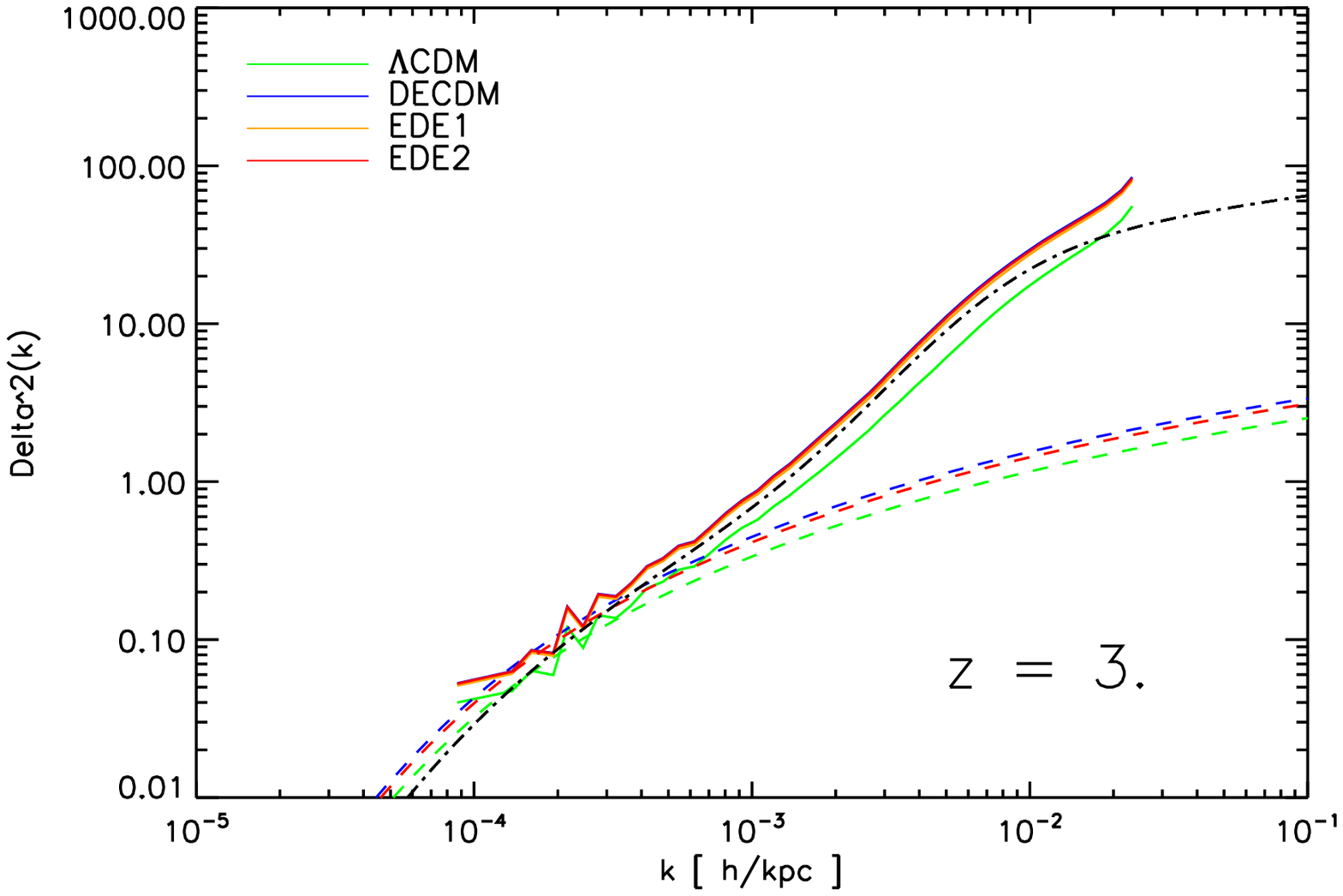}\\
\includegraphics[width=0.45\textwidth]{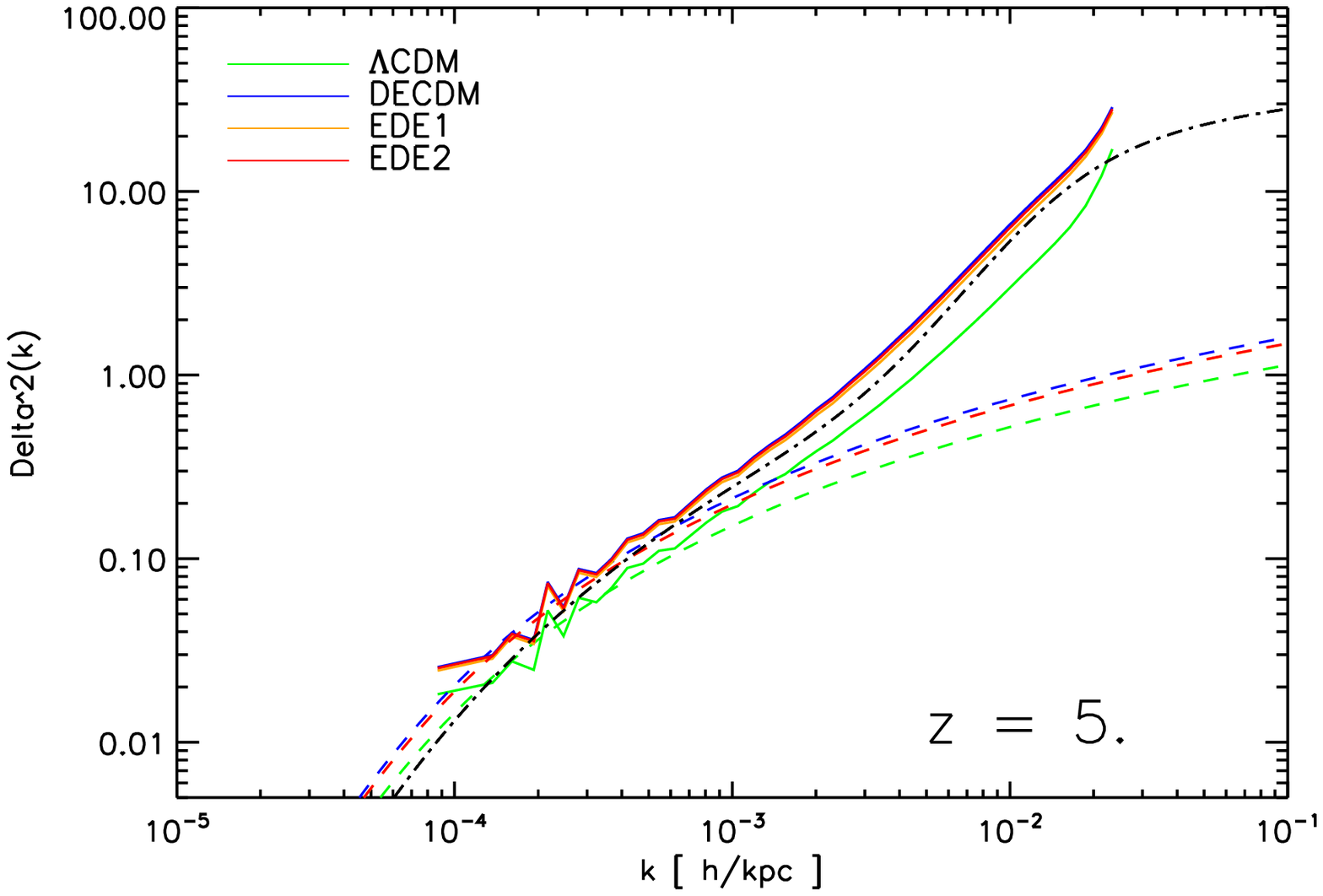}
\caption{Comparison of the non-linear power spectra of the four different
  cosmological models studied here. The three panels give results for
  redshifts $z=0$, $z=3$ and $z=5$, from top to bottom. The $y$-axes shows the
  dimensionless power $\Delta^{2}=k^3P(k)$ as a function of $k$ computed from
  the dark matter density field using a grid of $512^3$ points. All the
  simulations are normalized to $\sigma_8=0.8$ for the linearly extrapolated
  density field today. The dashed lines indicate the expected linear
  power spectra. The prediction from \citet{Smith2003} for the $\Lambda$CDM cosmology is shown by the black dot-dashed line.}
\label{fig:power}
\end{figure}

We can easily see that for very large redshift we recover the matter dominated
behaviour in the $ \Lambda$CDM case: $D(a) \propto a$.  On the other hand, as
expected, the linear growth in the two EDE models falls behind the green curve
in Fig.~\ref{fig:grow}, implying that they reach a given amplitude at earlier
times. In fact, the expansion rate in the $\Lambda$CDM cosmology is lower than
in EDE models, which governs the friction term $\left(\dot{a}/a\right)$ in the
growth equation
\begin{equation}
	\ddot{\delta}+2\frac{\dot{a}}{a}\dot{\delta}-4\pi G \rho \delta = 0
\end{equation}
of the perturbations.

These formulae can be used to derive a suitable expression for the reduced
linear overdensity $\delta_c$ for collapse expected in EDE models
\citep{Bartelmann2006}, which in turn suggests that there are significant
consequences for the process of non-linear structure formation, an expectation
that we will analyse later in detail.  In sum, structures need to grow earlier
in EDE models than in $\Lambda$CDM in order to reach the same amplitude at the
present time. At an equal redshift, the initial conditions must hence be more
evolved in order to produce comparable results today. The DECDM shows a
behaviour qualitatively similar to EDE1 and EDE2 (blue long-dashed line).


In all our simulations, we have identified dark matter halos using two
methods: the friends-of-friends (FOF) algorithm with linking length $b=0.2$,
and the spherical overdensity (SO) group finder.  Candidate groups with a
minimum of 32 particles were retained by the FOF group finder.  In the SO
algorithm, we first identify FOF groups, and then select the particle with the
minimum gravitational potential as their centres, around which spheres are
grown that enclose a fixed prescribed mean density $\Delta \times \rho_{\rm
  crit}$, where $\rho_{\rm crit}$ is the critical density. Different
definitions of virial overdensity are in use in the literature, and we
consider different values for $\Delta$ where appropriate.  The classical
definition of NFW adopts $\Delta = 200$ independent of cosmology, while
sometimes also $\Delta = 200\, \Omega_m$ is used, corresponding to a fixed
overdensity relative to the background density. Finally, a value of
$\Delta\sim 178\, \Omega_m^{0.45}$ based on a generalization of the spherical
top-hat collapse model to low density cosmologies can also be used. Note
however that this may in principle depend on the dark energy cosmology
\citep{Bartelmann2006}, and is hence slightly ambiguous in these cosmologies.

We have verified the correctness of our implementation of early dark energy in
the simulation code by checking that it accurately reproduces the expected
linear growth rate in these non-standard cosmologies.  Recall that rather than
normalizing the density perturbations of the initial conditions to the same
value at the (high) starting redshift, we determine them such that they should
grow to the same linear amplitude today in all of the models. In practice, we
fix $\sigma_8$, the linearly extrapolated {\it rms} fluctuations in top-hat
spheres of radius $8\,h^{-1}{\rm Mpc}$ to the value $0.8$ for the epoch $z=0$.

In Figure~\ref{fig:power}, we show measurements of the power spectrum of our
different models at three different redshifts. While the models differ
significantly at high redshift, the four different realizations show the same
amplitude of the power spectrum, at least on large scales, at redshift $z=0$
(left panel).  The fluctuations on small scales probably reflect the earlier
structure formation time in the EDE models and the resulting differences in
the non-linear halo structures. The good agreement of the power spectrum at
the end, as well as a detailed comparison of the growth rate of the largest
modes in the box with the linear theory expectation (not shown), demonstrate
explicitly that the EDE models are simulated accurately by the code, as
intended.

Note also that the power spectrum measurements show that due to the slower
evolution of the linear growth factor in EDE models, the degeneracy between
the models is lifted towards high-z since this corresponds to more time for
the different growth dynamics to take effect. Consequently, we expect a
different evolution of structures back in time. Our main focus in the
following will be to study the impact of a different equation of state for the
dark energy upon the mass function of dark matter halos and its evolution with
redshift.

\begin{figure*}
\centering
\includegraphics[width=0.33\textwidth]{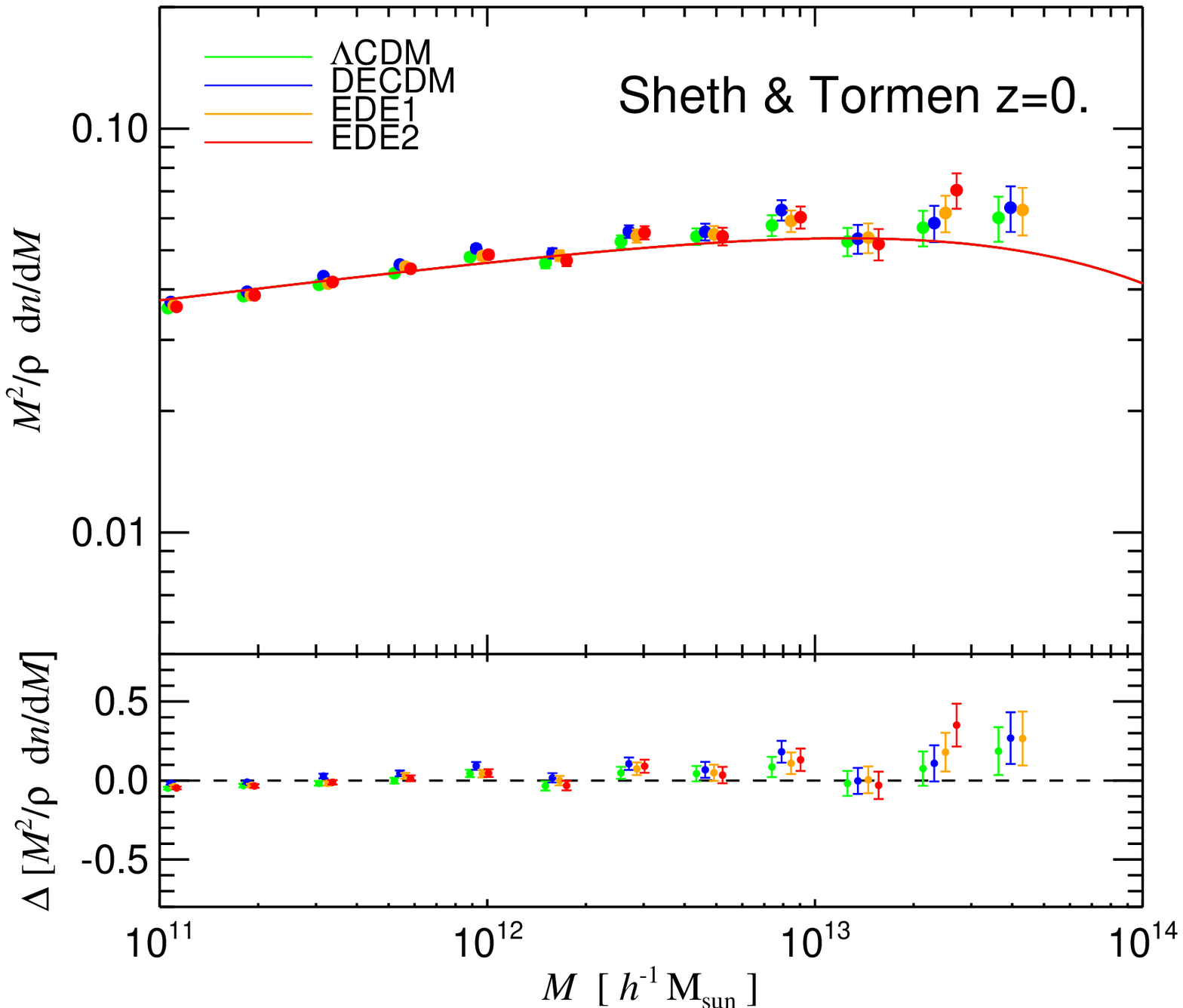}
\hfill
\includegraphics[width=0.33\textwidth]{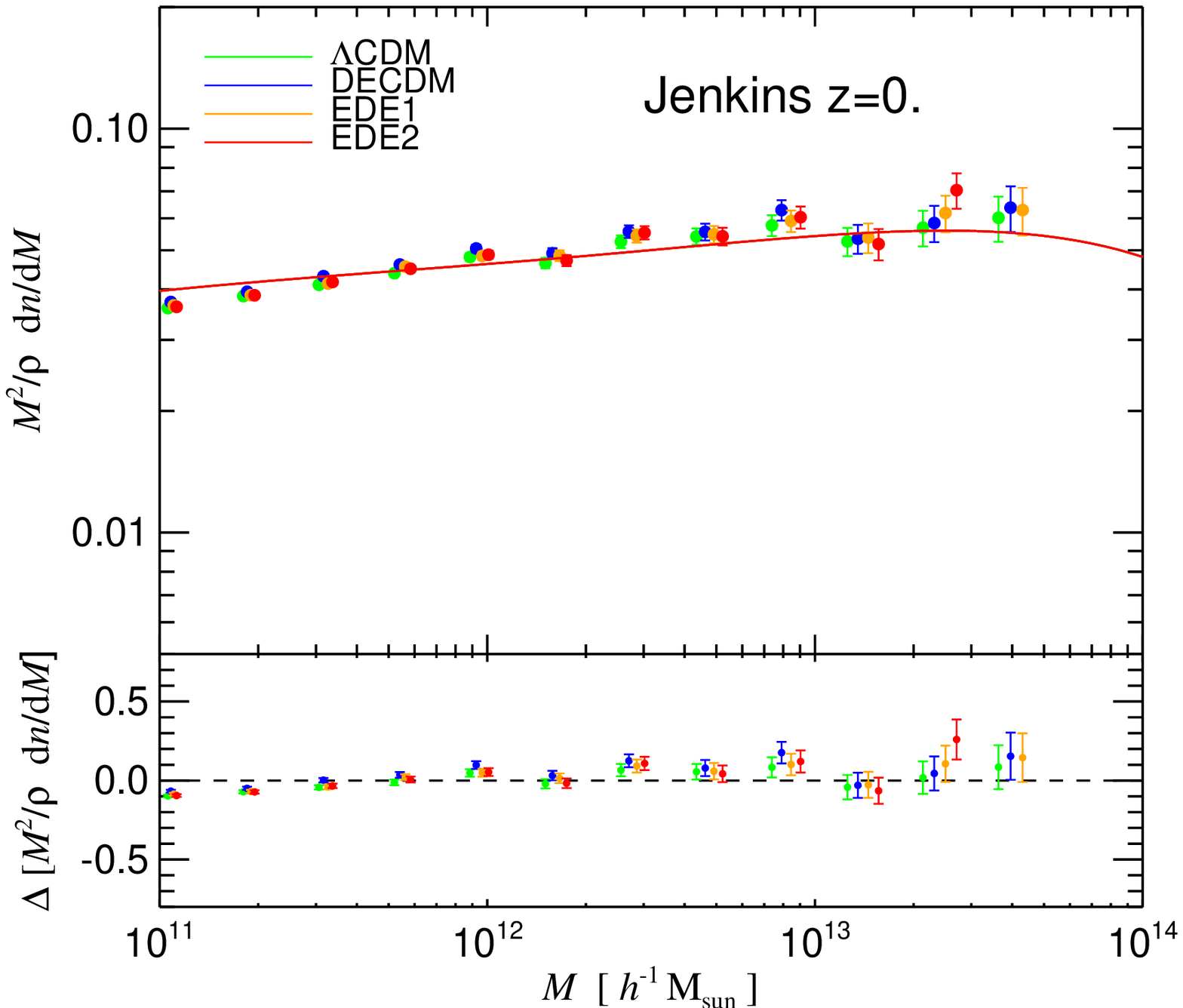}
\hfill
\includegraphics[width=0.33\textwidth]{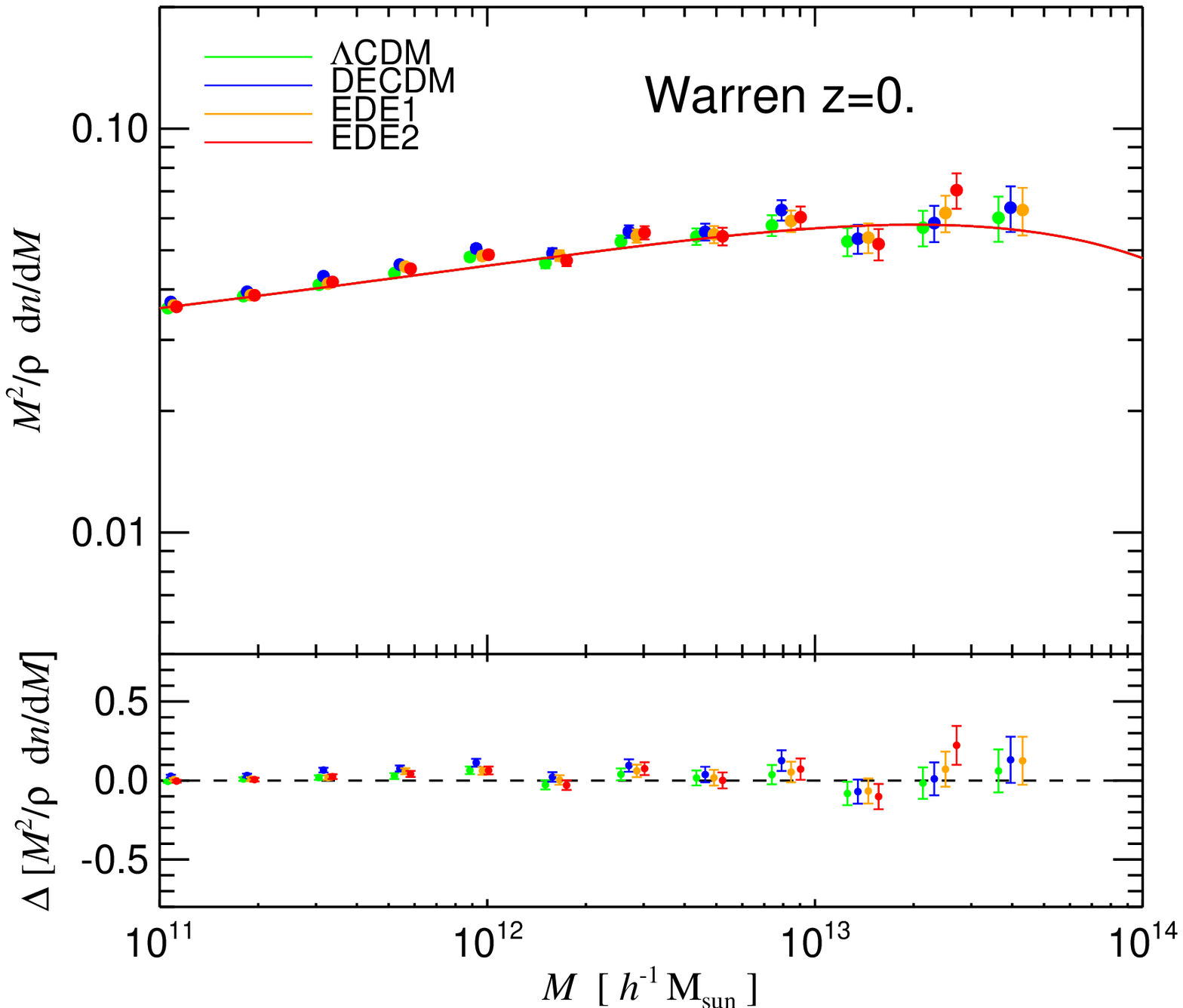}
\caption{Friends-of-friends multiplicity mass functions at $z=0$ for the four
  dark energy models studied here. The solid lines in each panel represent the
  multiplicity function computed analytically either from the Sheth \& Tormen
  formula (left panel), the Jenkins formula (central panel) or the Warren
  model (right panel). The symbols are the numerical simulation results for
  $\Lambda$CDM (green), DECDM (blue), EDE1 (orange) and EDE2 (red). We
  consider only halos with more than $200$ particles and we apply an upper
  mass cut-off where the Poisson error reaches $14\%$.  In the lower plot of
  each figure we show the residuals between analytically expected and
  numerically determined mass functions for all models. The differences are
  typically below $10\%$. The error bars show Poisson uncertainties due to
  counting statistics for all models. At $z=0$, the simulation results for all
  cosmologies are basically identical, which reflects the fact that we
  normalized the models such that they have the same linear power spectra
  today, with a normalization of $\sigma_{8}=0.8$. }
\label{fig:massfunctionz0}
\end{figure*}

\section{The mass function}

\begin{figure*}
\centering
\includegraphics[width=0.48\textwidth]{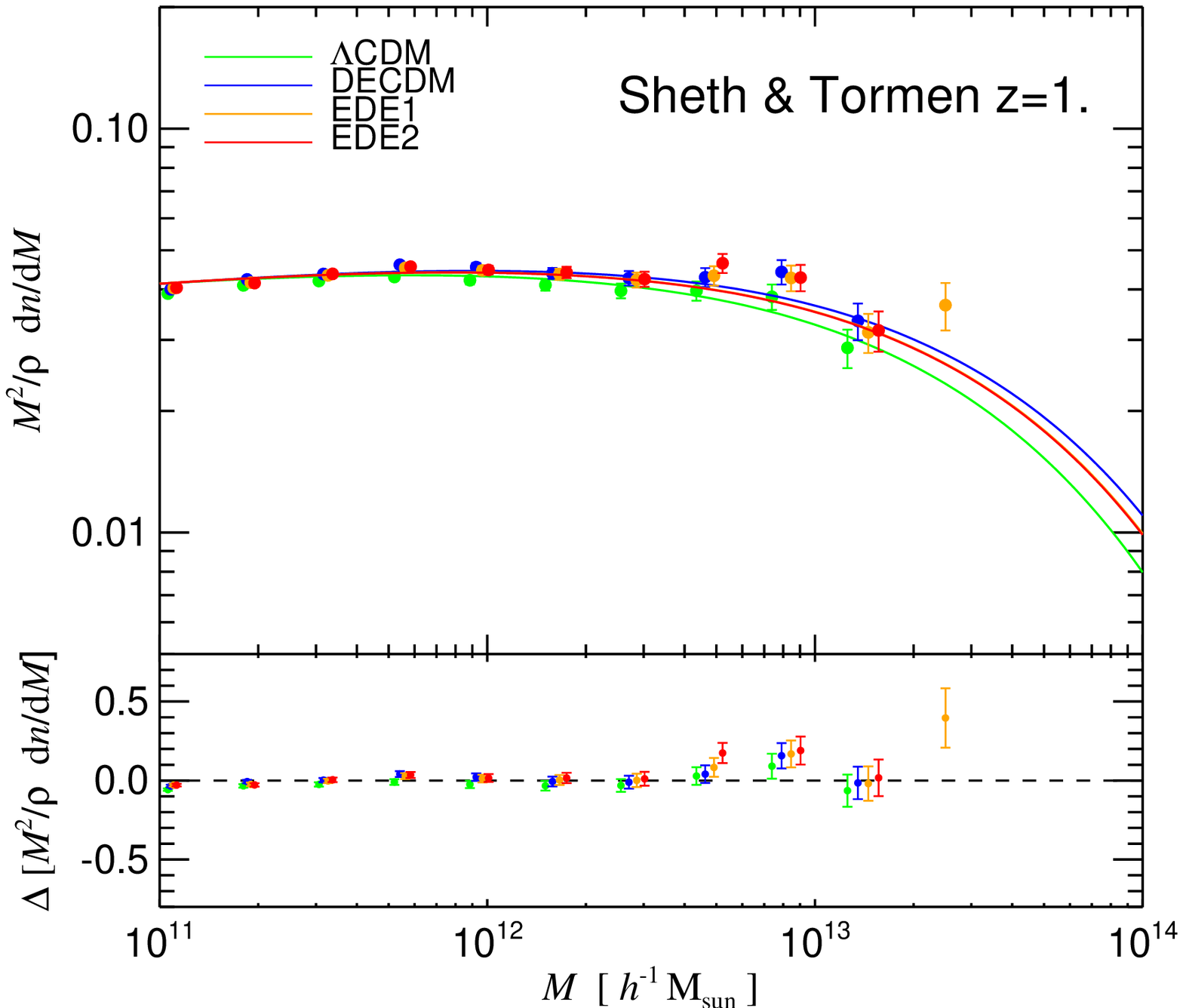}
\hfill
\includegraphics[width=0.48\textwidth]{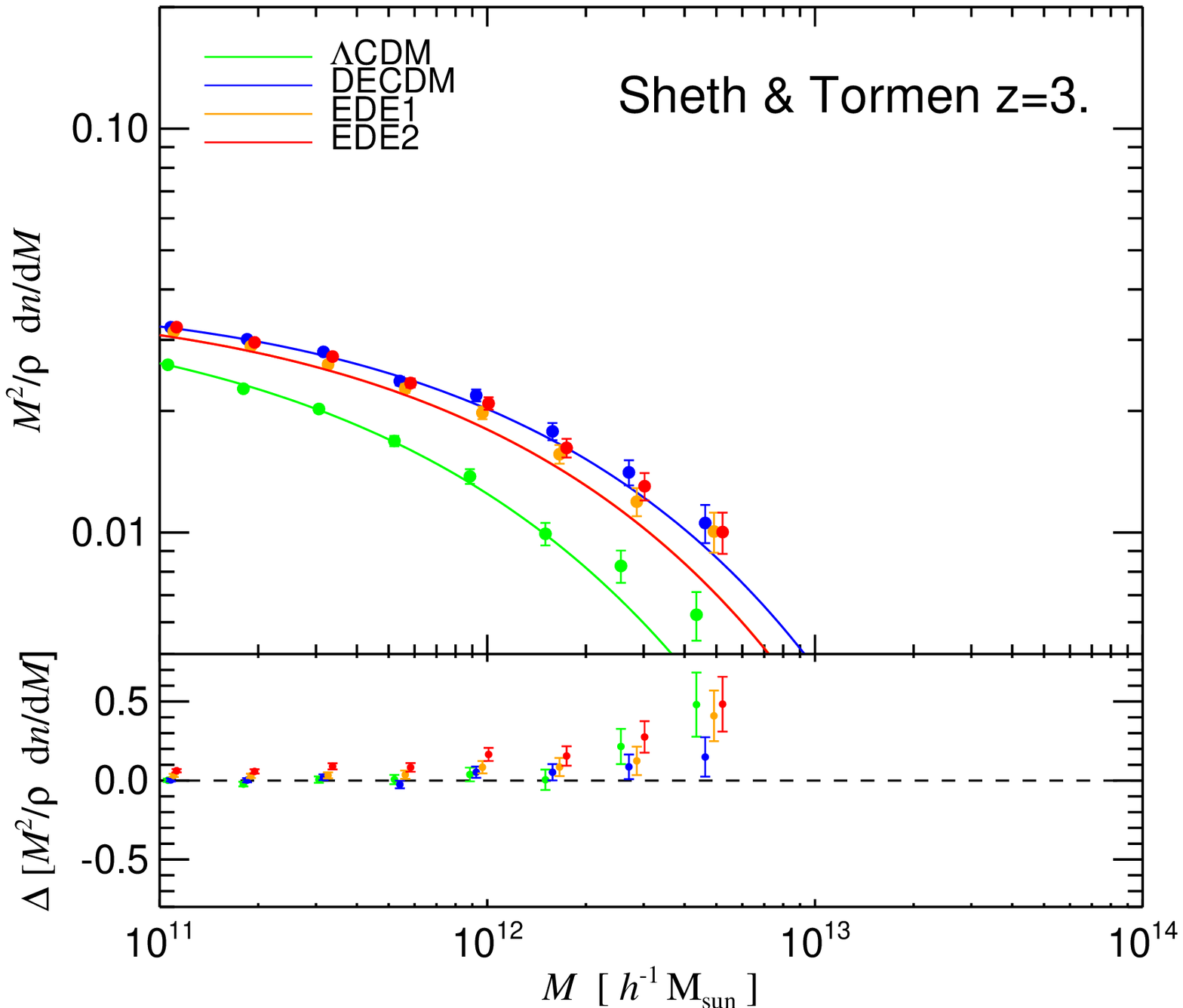}
\vskip 0.2 cm
\centering
\includegraphics[width=0.48\textwidth]{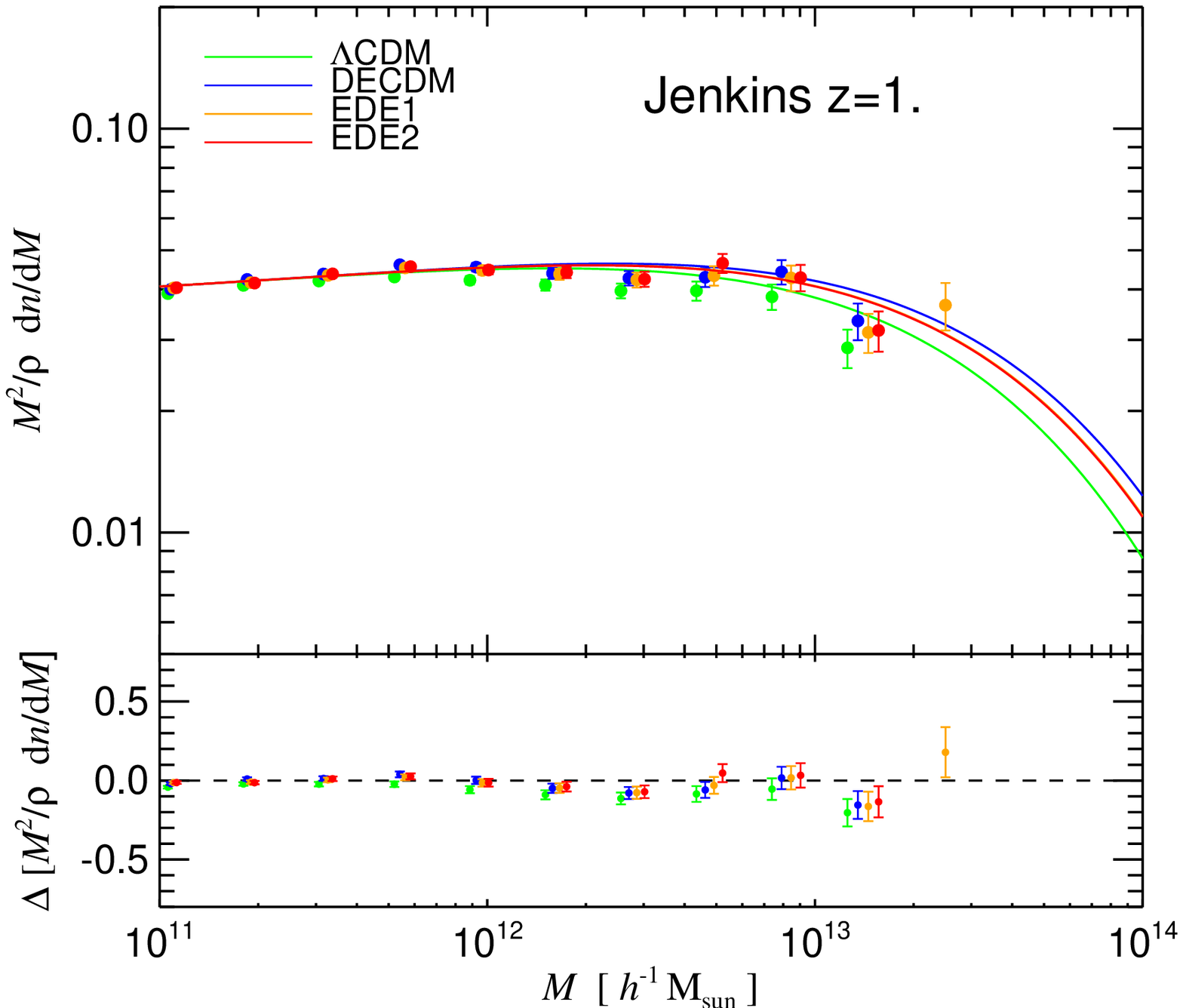}
\hfill
\includegraphics[width=0.48\textwidth]{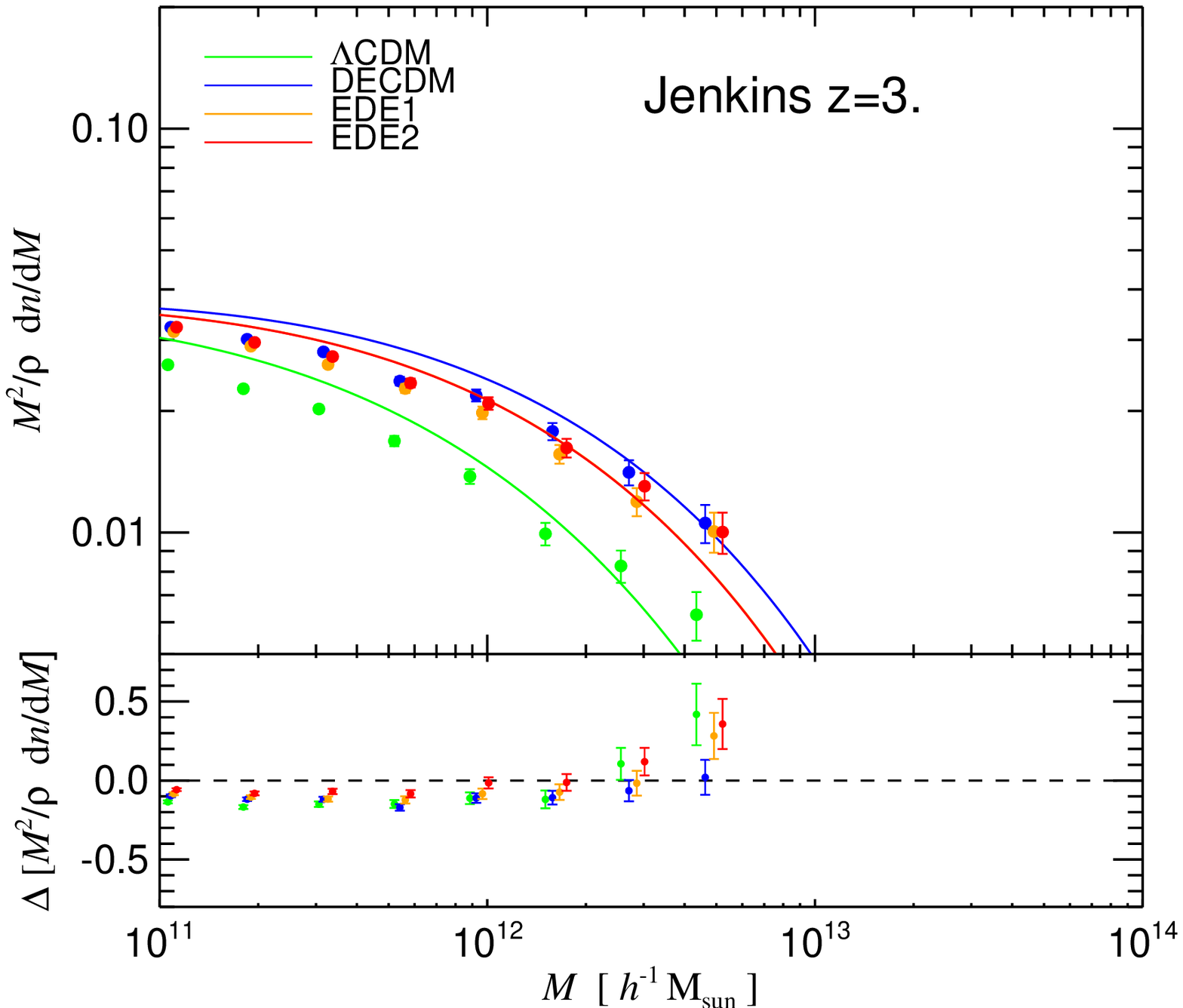}
\vskip 0.2 cm
\centering
\includegraphics[width=0.48\textwidth]{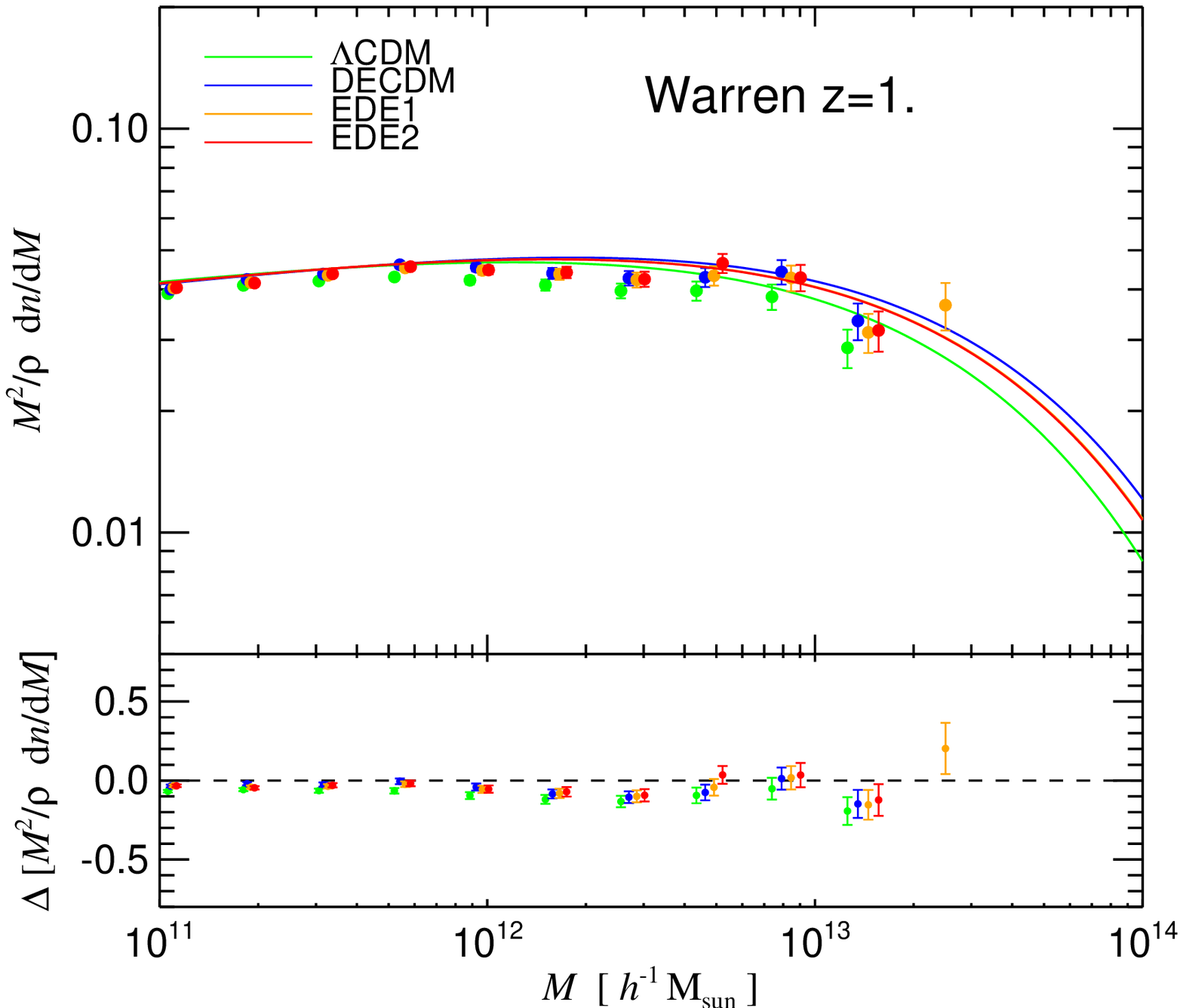}
\hfill
\includegraphics[width=0.48\textwidth]{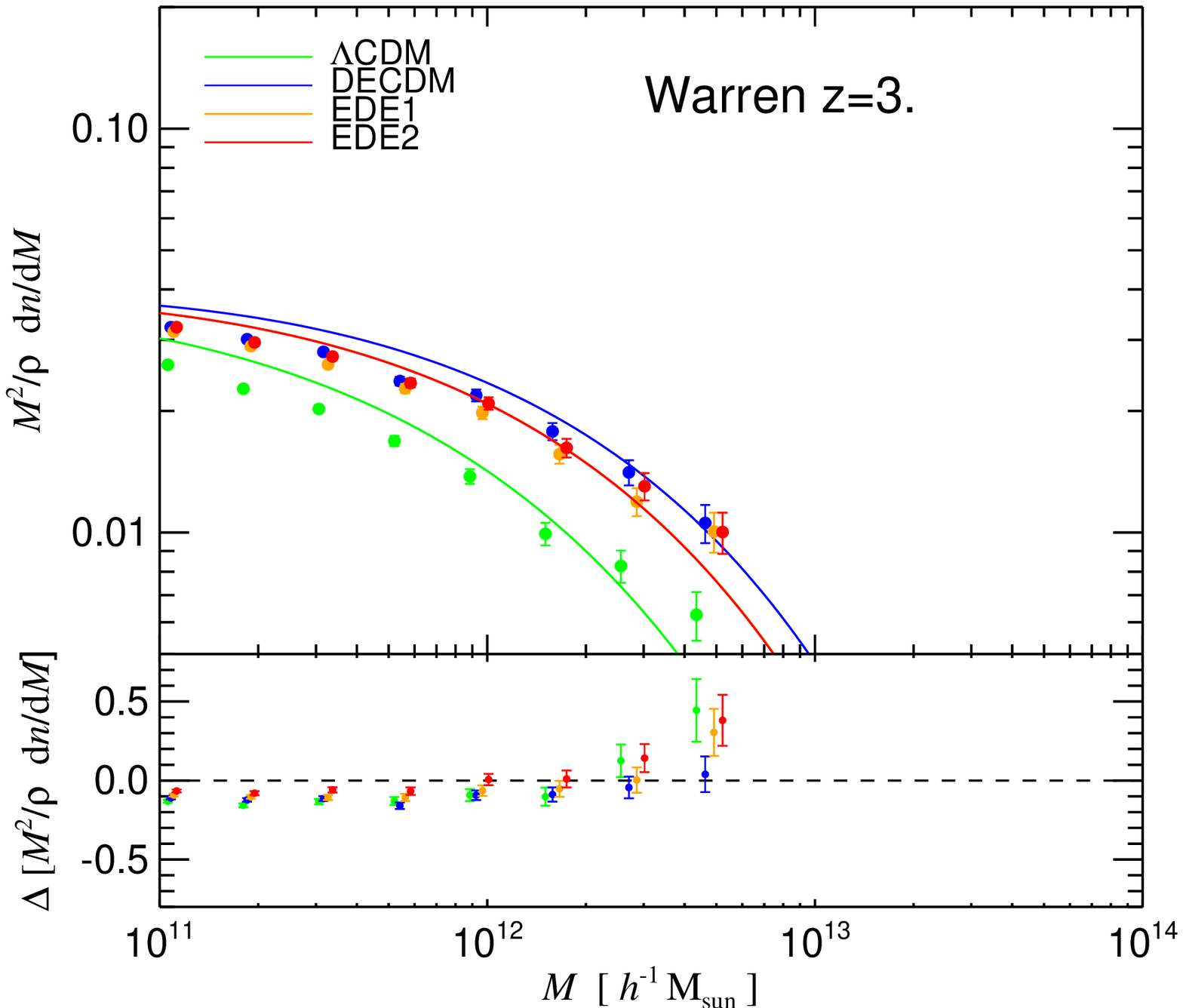}
\caption{Friends-of-friends multiplicity mass functions for the four dark
  energy models studied here. The evolution towards high redshift is shown in
  terms of results at $z=1$ (left column) and at $z=3$ (right column).  The
  solid lines in each plot represent the multiplicity function computed
  analytically from the Sheth \& Tormen formula (top row), the Jenkins formula
  (middle row) and the Warren formula (bottom row). The points are the
  numerical simulation results for $\Lambda$CDM model (green), DECDM (blue),
  EDE1 (orange) and EDE2 (red). We consider only halos with more than $200$
  particles and we apply an upper mass cut-off where the Poisson error reaches
  $14\%$.  In the lower plot of each figure we show the residuals between
  analytically expected and numerically determined mass functions for all
  models. The differences are typically below $15\%$. The error bars show
  Poisson uncertainties due to counting statistics for all models.}
\label{fig:massfunction}
\end{figure*}

In this section we measure the halo abundance at different redshifts and
compare with analytic fitting functions proposed in the literature. Our
primary goal is to see to which extent dark energy models can still be
described by these fitting formulae, and whether there is any numerical
evidence that supports the higher halo abundance predicted for the EDE
cosmologies \citep{Bartelmann2006}. We will mostly focus on halo mass
functions determined with the FOF algorithm with a linking length of 0.2, but
we shall also consider SO mass functions later on.

In Figure \ref{fig:massfunction}, we show our measured halo mass functions in
terms of the multiplicity function, which we define as
\begin{equation}
f\left(\sigma,z\right)=\frac{M}{\rho_{0}}\frac{{\rm d}n(M,z)}{{\rm d}\ln \sigma^{-1}}
\end{equation}
where $\rho_{0}$ is the background density, $n(M,z)$ is the abundance of halos
with mass less than $M$ at redshift $z$, and $\sigma$ is the mass variance of
the power spectrum filtered with a top-hat mass scale equal to $M$.  We give
results for the cosmological models $\Lambda$CDM, DECDM, EDE1 and EDE2,
plotted as symbols, while the solid lines show various theoretical
predictions.

Note that we plot the mass function only in a limited mass range in order to
avoid being dominated by counting statistics or resolution effects. To this
end we only consider halos above a minimum size of 200 particles. At the high
mass end, individual objects are resolved well, but the finite volume of the
box limits the number of massive rare halos we can detect. We therefore plot
the mass function only up to the point where the Poisson error reaches $\sim
14 \%$ (corresponding to minimum number of $\sim 50$ objects per bin).

As is well known, the Press \& Schechter mass function \citep{PS1974}, while
qualitatively correct, disagrees in detail with the results of N-body
simulations \citep{Efstathiou1988,White1993,Lacey1994,Eke1996}, specifically,
the PS formula overestimates the abundance of halos near the characteristic
mass $M_\star$ and underestimates the abundance in the high-mass tail.  We
therefore omit it in our comparison.  The discrepancy is largely resolved by
replacing the spherical collapse model of the standard Press \& Schechter
theory with the refined ellipsoidal collapse model
\citep{ST1999,ST2001,ST2002}. Indeed, in the top left panel of
Figure~\ref{fig:massfunctionz0} we can see quite good agreement of the Sheth
\& Tormen mass function (ST) with our simulations at $z=0$. We stress that
here the standard value of $\delta_c=1.689$ for the linear collapse threshold
has been used irrespective of the cosmological model.  Two other well-known
fitting formulae are that from Jenkins \citep[central
panel,][`J']{Jenkins2001} and that from Warren \citep[right
panel,][`W']{Warren2006}, which differ only very slightly in the low-mass
range.  We compare our measurements with these models in the panels of the
middle and right columns.  As we can see from the comparison between the solid
lines and the numerical data points, the differences between the different
theoretical models (which only rely on the linearly evolved power spectrum at
each epoch) and the simulation results is very small.

Figure~\ref{fig:massfunction} shows the redshift evolution of the mass
function, in the form of separate comparison panels at redshifts $z=1$ and
$z=3$. While at $z=0$ the different cosmologies agree rather well with each
other, as expected based on the identical linear power spectra, at redshift
$z=1$ we begin to see differences between the models, and finally at $z=3$, we
can observe a significantly higher number density of groups and clusters in
the non-standard dark energy models. Notice that the model with constant $w$
(blue line) behaves qualitatively rather similar to the EDE models. In each of
the panels, we include a separate plot of the residuals with respect to the
analytic fitting functions.  This shows that at $z=3$ the agreement is clearly
best for the ST formula.

The differences between the models are most evident in the exponential tail of
the mass function where it begins to fall off quite steeply, in agreement with
what is expected from the power spectrum analysis.  We can see that, at
high-$z$, replacing the cosmological constant by an early dark energy scenario
has a strong impact on the history of structure formation. In particular,
non-linear structures form substantially earlier in such a model, such that a
difference in abundance of a factor of $\sim 2$ is reached already by $z=3$.
This underlines the promise high redshift cluster surveys hold for
distinguishing different cosmological models, and in particular for
constraining the dynamical evolution of dark energy.

We now want to assess in a more quantitative fashion the differences between
our numerical halo mass functions and the analytic fitting functions. In
particular we are interested in the question whether we can objectively
determine a preference for one of the analytic models, and whether there is
any evidence that the ordinary mass function formalism does work worse for the
generalized dark energy models than for $\Lambda$CDM. The latter would
indicate that the critical linear overdensity threshold $\delta_c$ needs to be
revised for EDE models, as suggested by the analytic spherical collapse theory
\citep{Bartelmann2006}.

\begin{figure}
\begin{center}
\includegraphics[width = 0.47\textwidth]{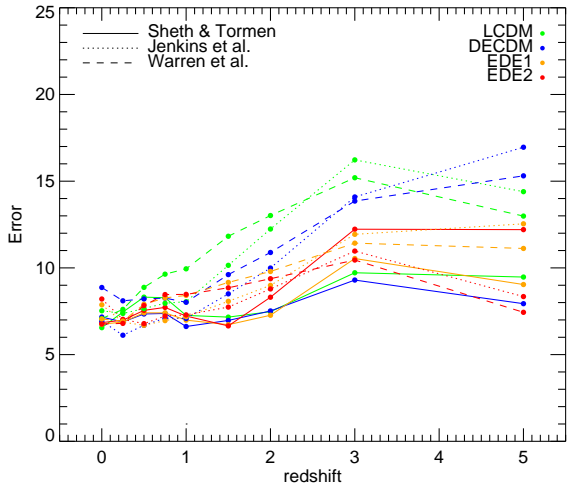}
\caption{Redshift dependence of our goodness of fit parameter $\chi^2$
  (see Eqn.~\ref{eq:chi2}), expressed as percent, computed by comparing the theoretical expectation for
  the multiplicity mass function with the simulation results. All cosmological
  models are compared. The deviations are computed with respect to the Sheth
  \& Tormen model (solid lines), the Jenkins et al. (dotted lines) and the
  Warren (dashed lines). }
\label{fig:error}
\end{center}
\end{figure}

To this end we directly measure the goodness of the fit, which we define for
the purposes of this analysis as:
\begin{equation}
  \chi^2= \left( \sum_j 1/\sigma_j^2\right)^{-1}  \sum_{i}\frac{({\rm MF}_{i}-{\rm
      MF}_{{\rm TH},i})^{2}}{ \sigma_i^{2} \, {\rm MF}_{{\rm TH},i}^2}, 
\label{eq:chi2}
\end{equation}
where ${\rm MF}_{{\rm TH},i}$ are the theoretical values, ${\rm MF}_{i}$ are
the simulations results, and we took into account a simple Poisson error in
the definition of the goodness of fit. In Figure~\ref{fig:error}, we plot this
value expressed in percent for all simulations when compared with the
theoretical formulae of ST (solid line), Jenkins (dotted line) and Warren
(dashed line).  We cannot identify a clearly superior behaviour of any the
three fitting functions, at least at this level of resolution; the models lie
in a strip between approximately $5$ and $15 \%$ error between $z=0$ and
$z=5$. There is some evidence that the ST model does a bit better than the
other fitting formulae for the $\Lambda$CDM cosmology at high redshift, but
the opposite is true for the two EDE cosmologies and the Jenkins and Warren
functions.

\begin{figure}
\begin{center}
\includegraphics[width = 0.47\textwidth]{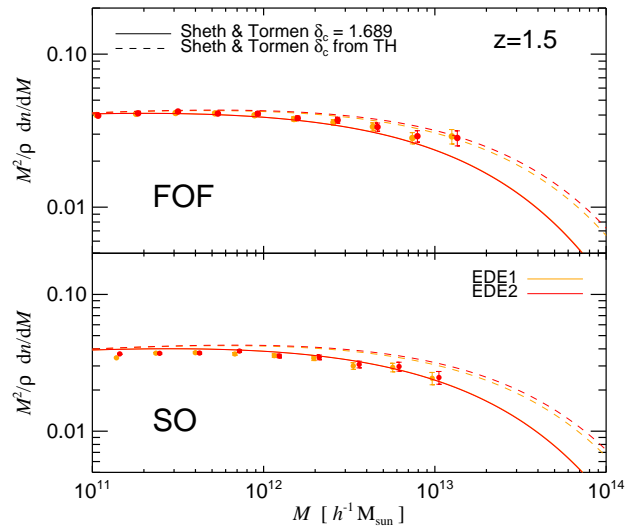}
\caption{Multiplicity mass function at $z=1.5$ for the two EDE models studied
  here. We want to highlight that the introduction of a modified overdensity
  motivated by the generalized spherical collapse theory
  \citep{Bartelmann2006} reduces the agreement between the simulation results
  and the theoretical ST mass function. The measured points are in better
  agreement with the solid line (standard ST model) than with the dashed lines
  (ST modified model), the latter systematically overestimate the halo
  abundance. This holds true both if we consider the halos obtained from the
  FOF halo finder (top panel) or the one obtained by taking into account the
  theoretically motivated virial overdensity with the appropriate SO mass
  definition (bottom panel). }
\label{fig:diffDelta}
\end{center}
\end{figure}

Interestingly, the overall agreement between simulation results and fitting
functions is actually slightly worse for $\Lambda$CDM than for the
non-standard dark energy cosmologies. There is hence no tangible evidence that
a revision of the mass function formalism is required to accurately describe
EDE cosmologies. Our finding of a universal $f(\sigma)$ is quantitatively
different from the expectation based on the analysis of the EDE models by
\citet{Bartelmann2006}.  We find that only the different linear growth rate
has to be taken into account for describing the mass function in the early
dark energy cosmologies with the ST formalism, but there is no need to modify
the linear critical overdensity value. To make this point more explicit, we
show in Figure~\ref{fig:diffDelta} the mass function for the EDE models and
compare it to standard ST (solid lines), and to the expectations obtained
taking into account a different density contrast for EDE models (dashed
lines). The predictions in the second case are based on the analytic study of
\citet{Bartelmann2006} and the critical overdensity is proportional to $(a)^{3
  \Omega_{\rm de,sf}/5}$ (see Eqn.~\ref{eq:Omegasf}).\footnote{In order to
  obtain the new values for the critical overdensity it is necessary to
  compute the virial overdensity by solving the equation of the generalized
  spherical collapse model.} Clearly, the proposed modification of $\delta_c$
actually worsens the agreement, both for the halos selected according the FOF
algorithm (top panel) or defined with respect to the virial overdensity
(bottom panel).

In the plots we discussed above, we always employed the FOF halo finder with
standard linking length of $b=0.2$ to find the halos, and the masses were
simply the FOF group masses, which effectively correspond to the mass within
an isodensity surface of constant overdensity relative to the background
density.  As the analytic mass function formulae have been calibrated with FOF
halo mass functions, we expect that they work best if the mass is defined in
this way.  However, we may alternatively also employ a different mass
definition based on the spherical overdensity (SO) approach, which allows one
to take into account the time-dependent virial overdensity $\Delta$ predicted
by generalizations of the spherical collapse model for dark energy
cosmologies. In the bottom panel of Figure~\ref{fig:diffDelta} we can see that
an even more marked disagreement results when we take into account this
arguably more consistent halo definition.
 
\begin{figure}
\begin{center}
\includegraphics[width=0.47\textwidth]{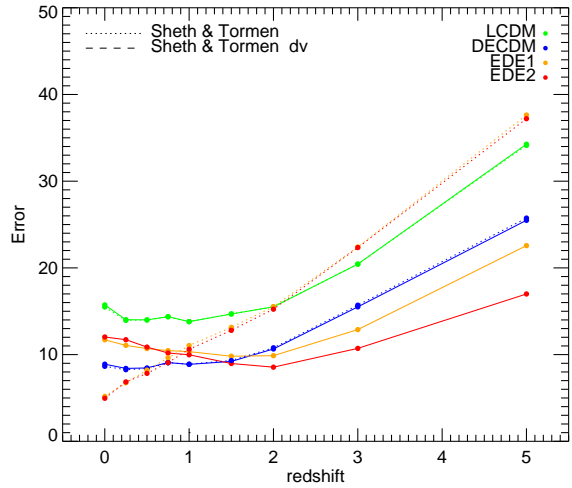}
\caption{Redshift dependence of our goodness of fit parameter $\chi^2$
  (see Eqn.~\ref{eq:chi2}), expressed as percent, computed by comparing the theoretical expectation for
  the multiplicity mass function with the simulation results. All models are
  considered. Here we use the top hat halo mass definition to compute the mass
  function from the simulations, and the deviations are computed with respect
  to the standard Sheth \& Tormen model (dotted-dashed lines), and the Sheth
  \& Tormen formula computed from a generalization of the top-hat collapse
  theory (dashed lines). }
\label{fig:errorTopHat}
\end {center}
\end{figure}

To stress this conclusion, in Figure \ref{fig:errorTopHat}, we show the
residuals of our SO halo mass functions compared with the Sheth \& Tormen
prediction, as a function of redshift and for our different cosmological
models, using the same procedure already applied to the FOF halo finder
results.  In this case, the halos were defined as virialized regions that are
overdense by a variable density threshold equal to
\begin{equation}
\Delta_c = 18\pi^{2} + 82x - 39x^{2},
\end{equation}
where $x=\Omega_{m}(z)-1$, see \cite{Bryan1998}. This is the predicted
dependence of $\Delta$ for $\Lambda$CDM, which we used for simplicity also for
the other dark energy cosmologies.  As expected, we see that the error
increases relative to the FOF mass functions, with discrepancies of order $10
\% $ at $z=0$.  However, there is again no evidence that the non-standard dark
energy cosmologies are described worse by the ST formalism than $\Lambda$CDM.
Also, there is no improvement in the accuracy of the fit when we introduce the
modified linear density contrast for the EDE models. On the contrary, as seen
by the dotted lines, which represent the theoretical mass function (based on
Sheth \& Tormen) modified according to the spherical top hat collapse theory
proposed by \cite{Bartelmann2006}.

Our results thus suggest that the mass function depends primarily on the
linear power spectrum and is only weakly, if at all, dependent on the details
of the expansion history. This disagrees with the expectations from the
generalization of the top hat collapse theory, which are not confirmed by our
numerical data.  In fact, our simulations show that a description of the mass
function based on the generalized TH calculation is incorrect at the accuracy
level reached here.  While the dynamic range of our results could be improved
by increasing the resolution and box-size of our simulations, it appears
unlikely that this could affect our basic conclusions.  Nevertheless, better
resolution would be required if one seeks to still further reduce the present
residuals of order 5-15\% between the fitting functions of ST, Jenkins or
Warren.

\begin{figure*}
\centering
\includegraphics[width=0.45\textwidth]{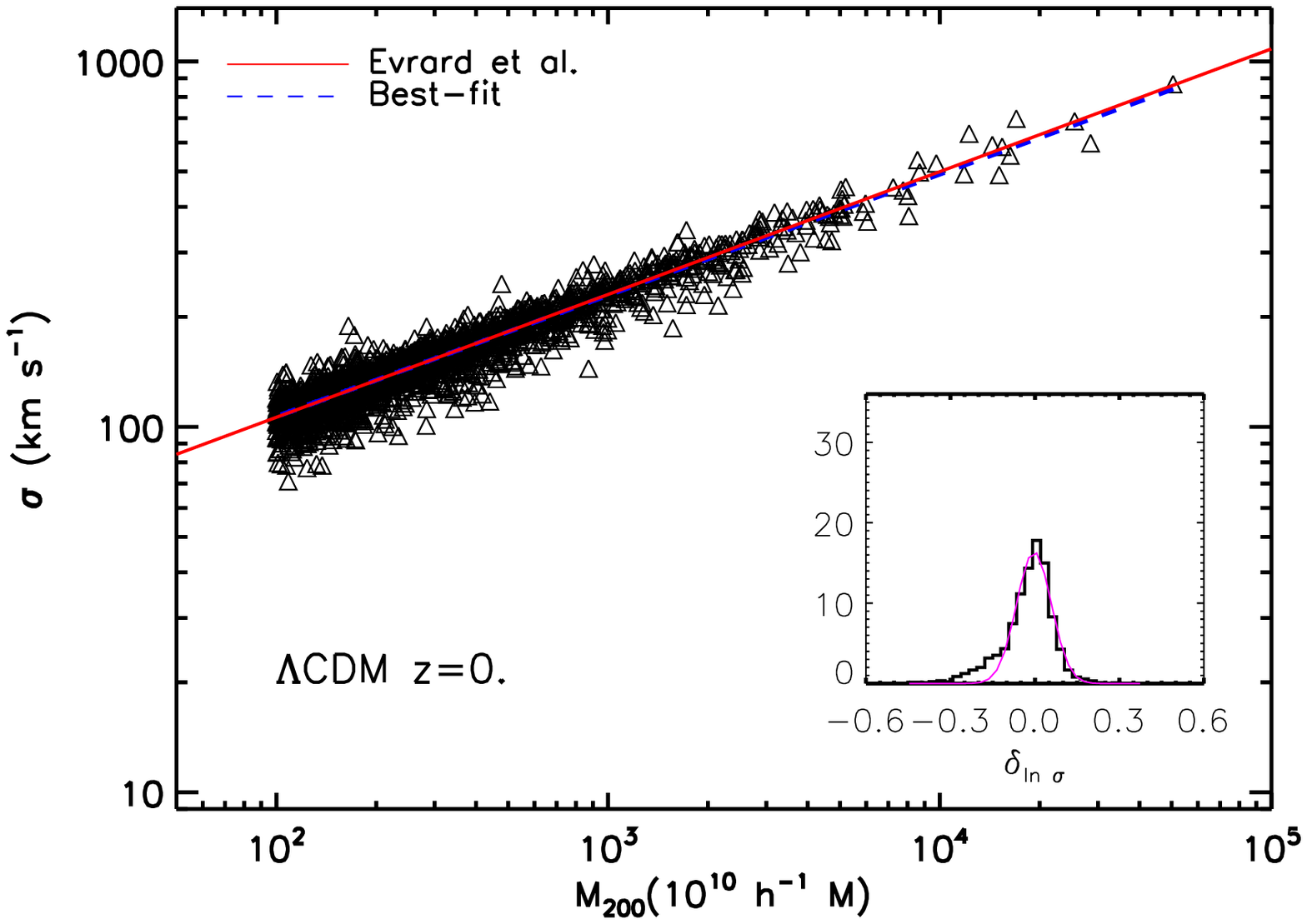}
\hfill
\includegraphics[width=0.45\textwidth]{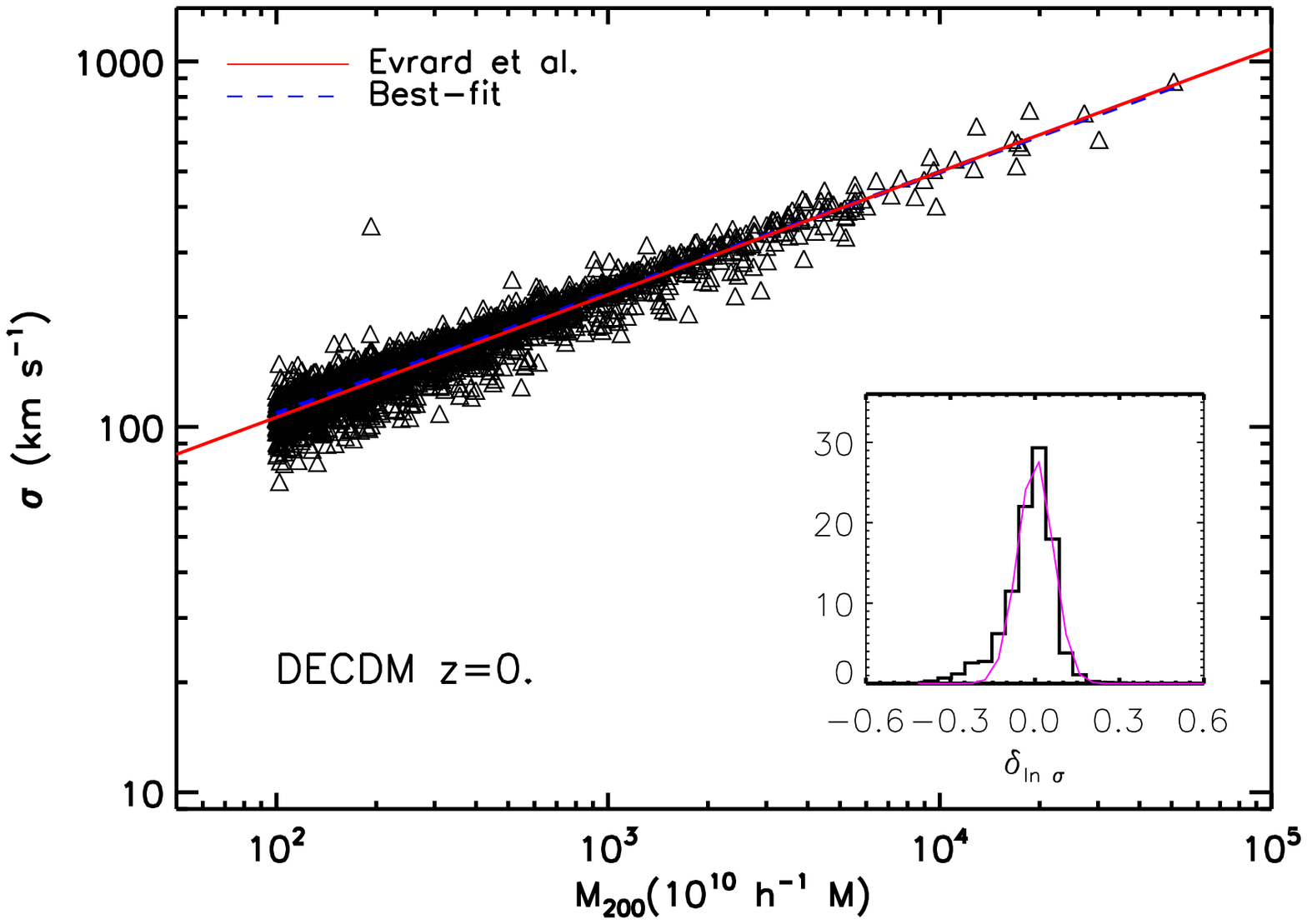}
\vskip 0.3 cm
\includegraphics[width=0.45\textwidth]{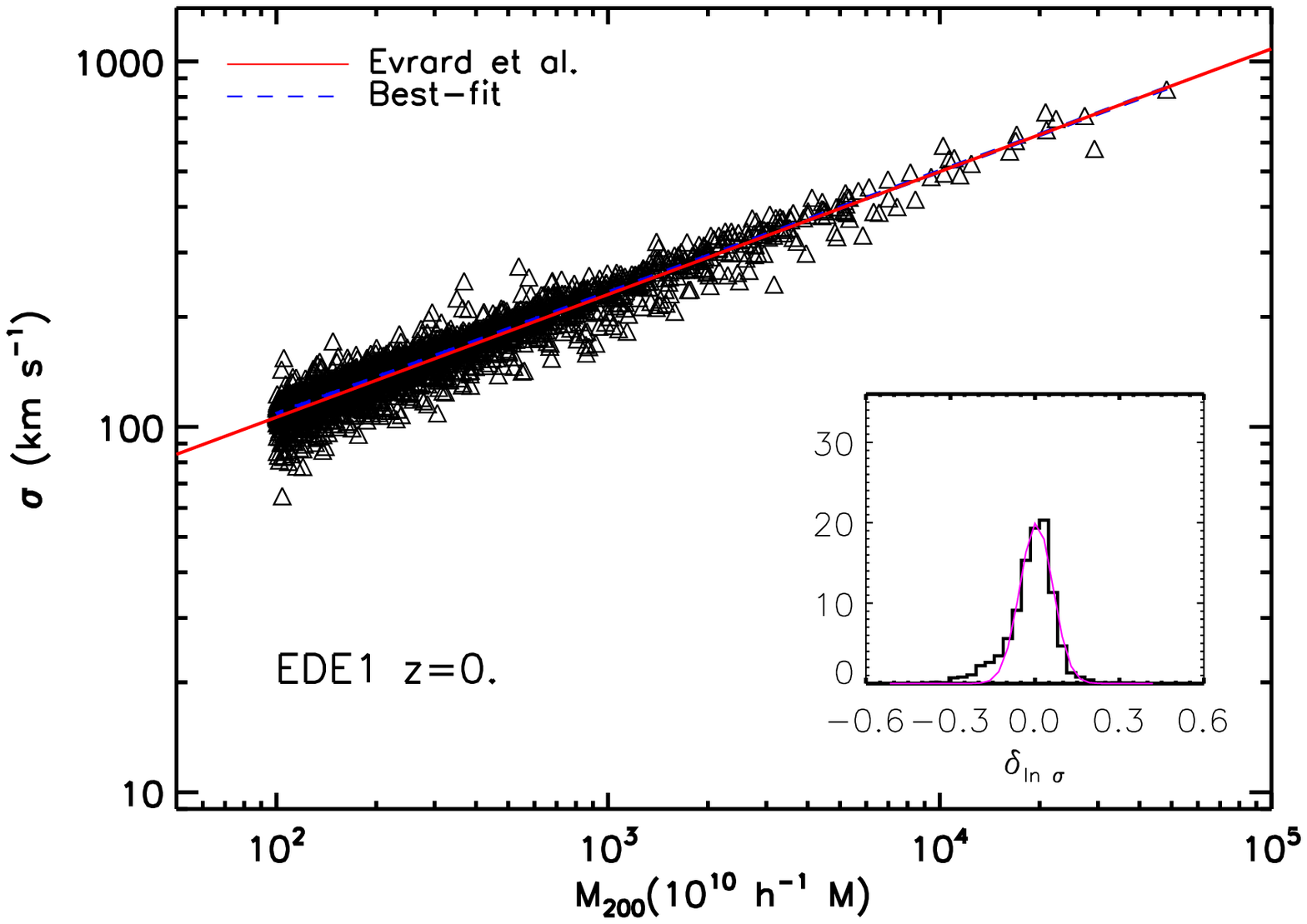}
\hfill
\includegraphics[width=0.45\textwidth]{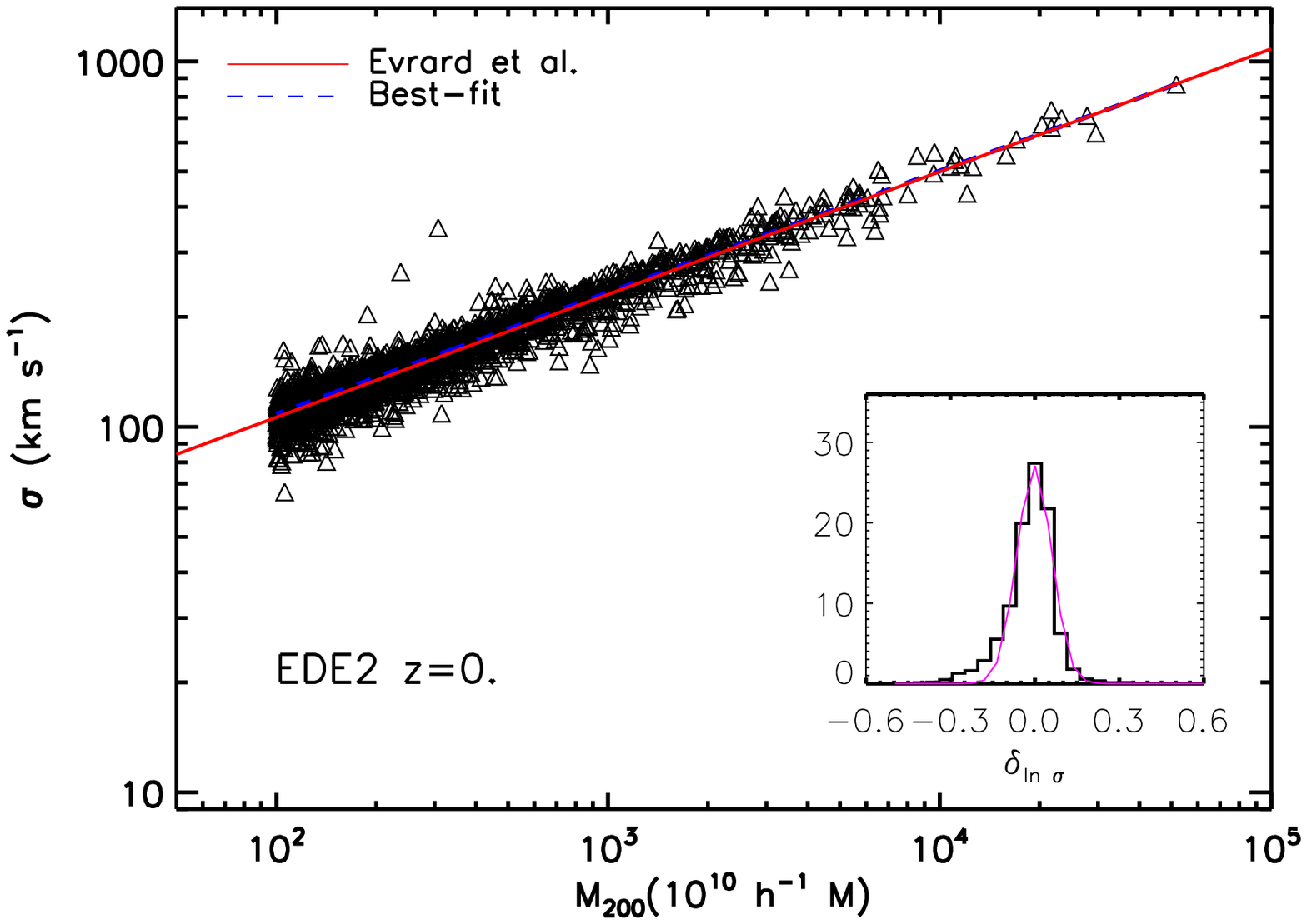}
\caption{The virial scaling relation at the present epoch for primary halos
  with mass larger than $10^{12}\,{\rm M}_{\odot}$ for the four models
  considered (from left to right and top to bottom: $\Lambda$CDM, DECDM, EDE1,
  EDE2). The red solid line in each plot represents the \citet{Evrard2008}
  relation, while the blue dashed line is our best fit. The triangles are the
  simulation results: we employ a fixed critical threshold of $\Delta = 200$
  to identify the dark matter halos.  The insets show the distributions of
  deviations in $\ln \sigma_{\rm DM}$ around the \citet{Evrard2008} fit. }
\label{fig:VelDisp}
\end{figure*}

\section{Halo properties}

\subsection{The virial scaling relation}

\cite{Evrard2008} have shown that the dark matter velocity dispersion of halos
provides for accurate mass estimates once the relationship between mass and
velocity dispersion is accurately calibrated with the help of numerical
simulations. They have demonstrated that there exists a quite tight power-law
relation between the mass of a halo and its one-dimensional velocity
dispersion $\sigma_{\rm DM}$, where
\begin{equation}	
\sigma_{\rm DM}^2=\frac{1}{3N_p}\sum^{N_p}_{i=1}\sum^{3}_{j=1}\left|v_{i,j}-\bar{v_{j}}\right|^{2},
\end{equation}
with $v_{i,j}$ being the $j_{\rm th}$ component of the physical velocity of
particle $i$ in the halo, $N_{p}$ is total number of halo particles within a
radius that encloses a mean overdensity of $\Delta=200$ with respect to the
critical density, and $ \bar{v}$ is the mean halo velocity.  When virial
equilibrium is satisfied, we expect that the specific thermal energy in a halo
of mass $M$ and of radius $R$ will scale with its potential energy, $GM/R$,
while the kinetic energy is proportional to $M^{2/3}$. Since $ \sigma_{\rm
  DM}$ expresses the specific thermal energy in dark matter, we can express
the mean expected velocity dispersion as a function of mass as
\begin{equation}
	\sigma_{\rm DM}\left(M,z\right) =
        \sigma_{\rm DM,15}\left(\frac{h(z)M_{200}}{10^{15}M_{\odot}}\right)^{\alpha}
        .
\end{equation}
Here the fit parameters are the slope $\alpha$ of the relation, and the
normalization $\sigma_{\rm DM,15}$ at a mass scale of $10^{15} h^{-1}
M_{\odot}$.  While the slope $\alpha$ just follows from the virial theorem if
halos form a roughly self-similar family of objects (which they do to good
approximation), the amplitude $\sigma_{\rm DM,15}$ of the relationship is a
non-trivial outcome of numerical simulations and reflects properties of the
virialization process of the halos as well as their internal
structure. \citet{Evrard2008} showed that a single fit is consistent with the
numerical data of a large set of N-body simulations of the $\Lambda$CDM
cosmology, covering a substantial dynamic range.

However it is conceivable that the amplitude of the relationship will be
slightly different in early dark energy cosmologies, as a result of the
different virial overdensity that is predicted by the top hat collapse in
these cosmologies. If true, this would then also hint at a different
normalization of the relationship between total Sunyaev-Zeldovich decrement
and mass, which would hence directly affect observationally accessible probes
of the cluster mass function at high redshift.

We here test whether we can find any difference in this relationship for our
different dark energy cosmologies.  In Figure~\ref{fig:VelDisp}, we plot the
velocity dispersion of halos as a function of mass, in the four different
cosmologies we simulated. The halos were identified using a spherical
overdensity definition, where the virial radius $r_{200}$ was determined as
the radius that encloses a fixed multiple of $200$ times the critical density
at the redshift $z$, and $M_{200}$ being the corresponding enclosed mass.  We
then determined the best-fit relation obtained from our numerical data (red
solid lines). This fit is in very good agreement with the results obtained by
\cite{Evrard2008} (dotted blue lines), given by $ \sigma_{\rm DM,15}=1082\pm
4.0\,{\rm km\, s^{-1}}$ and $ \alpha=0.3361\pm0.0026$, a value consistent with
the viral expectation of $\alpha=1/3$.  The insets show the residuals about
the fit at redshift $z=0$. They have a log-normal distribution with a maximum
of $6 \% $ dispersion (for the DECDM model) around the power-law relation. The
histograms are well fit by a log-normal with zero mean.

We find that the halos closely follow a single virial relation, insensitive of
the cosmological parameters, the epoch and also the resolution of the
simulation.  In particular, we do not find any significant differences for the
EDE models, instead, the same form of the virial relation is preserved across
the entire range of mass and redshift in the four simulations. The velocity
dispersion-mass correlation hence appears to be global and very robust
property of dark matter halos which is not affected by different contributions
of dark energy to the total energy density of the universe.

This is a reassuring result as it means that also in the case of early dark
energy, clusters can be studied as a one parameter family and the calibration
of dynamical mass estimates from internal cluster dynamics does not need to be
changed. Differences in the normalization should only reflect more or less
frequent halo mergers and interactions, which can introduce an additional
velocity component \citep{Ragone2007,Faltenbacher2007}.

\begin{figure*}
\begin{center}
\includegraphics[width = 0.8\textwidth]{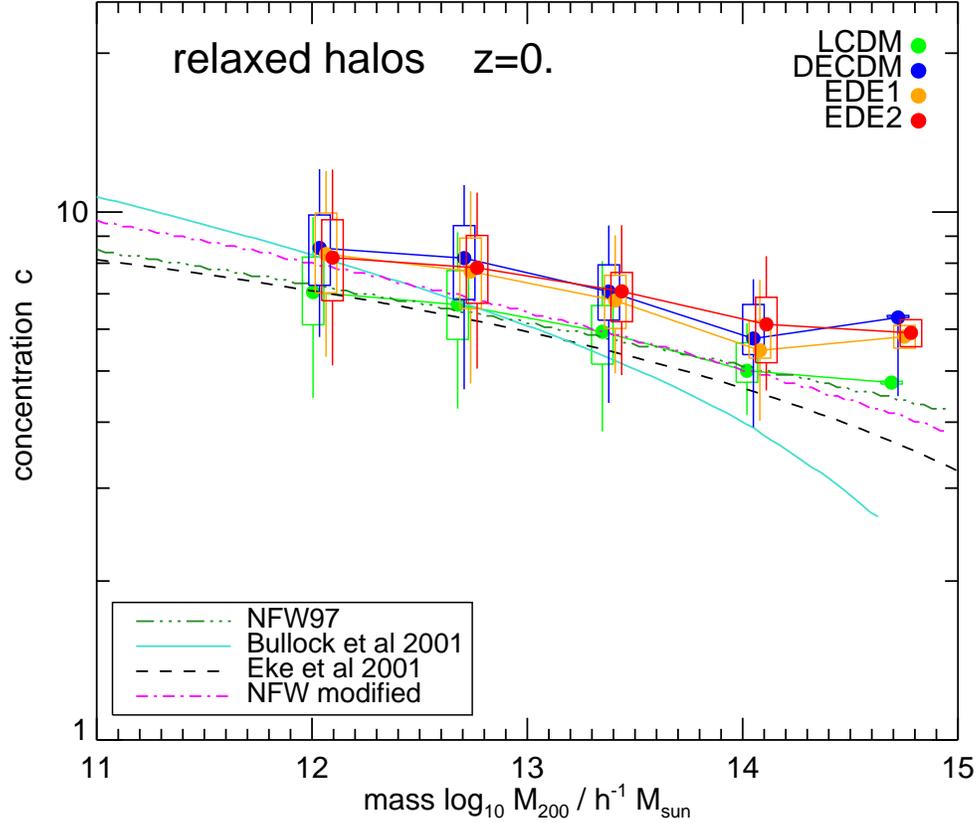}
\caption{Mass-concentration relation for relaxed halos in all our
  simulations. The boxes represent the 25 and 75 percentiles of the
  distribution with respect to the median value, while the whiskers show the 5
  and 95 percentiles. We compare our results with the theoretical expectations
  from NFW, ENS, B01. Also, a modified NFW prescriptions with slightly modified
  parameters as updated by \citet{Gao2008} is  shown (see Section~5 for
  details).}
\label{fig:concentration}
\end{center}
\end{figure*}

\subsection{Halo concentrations}

As we have seen, for an equal normalization of the present-day linear power
spectrum, the dark matter halo mass function at $z=0$ does not depend on the
nature of dark energy. On one hand this is a welcome feature, as it simplifies
using the evolution of the mass function to probe the expansion history of the
universe, but on the other hand it disappointingly does not provide for an
easy handle to tell different evolutions apart based only on the present-day
data.  However, a discrimination between the models may still be made if the
{\em internal structure} of halos is affected by the formation history, which
would show up for example in their concentration distribution.

Cosmological simulations have consistently shown that the spherically averaged
mass density profile of equilibrium dark matter halos are approximately
universal in shape. As a result, we can describe the halo profiles by the NFW
formula \citep{NFW1995,NFW1996,NFW1997}:
\begin{equation}
	\frac{\rho\left(r\right)}{\rho_{\rm crit}}=\frac{\delta_c}{(r/r_{s})(1+r/r_{s})^{2}},
\end{equation}
where $ \rho_{\rm crit} = 3H_0^2/8 \pi G$ is the critical density, $ \delta_c$
is the characteristic density contrast and $r_s$ is the scale radius of the
halo. The concentration $c$ is defined as the ratio between $r_{200}$ and
$r_s$. The quantities $\delta_c$ and $c$ are directly related by
\begin{equation}
  \delta_c=\frac{200}{3} \frac{c^3}{[\ln(1+c)-c/(1+c)]}.
\label{eq:deltacon}
\end{equation}

The concentration $c$ is the only free parameter in Eqn.~(\ref{eq:deltacon})
at a given halo mass and these two quantities are known to be correlated. In
fact, characteristic halo densities reflect the density of the universe at the
time the halos formed; the later a halo is assembled, the lower is its average
concentration.

We have measured concentrations for our halos in the different cosmologies
using the same procedures as applied to the analysis of the Millennium
simulation \citep{Neto2007,Gao2008}. For our measurements, we take into
account both relaxed and unrelaxed halos. In the second case, the equilibrium
state is assessed by means of three criteria: (1) the fraction of mass in
substructures with centers inside the virial radius is small, $f_{\rm
  sub}<0.1$, (2) the normalized offset between the center-of-mass of the halo
$\vec{r}_{\rm cm}$ and the potential minimum $\vec{r}_c$ is small, $s= |
\vec{r}_c-\vec{r}_{\rm cm} | /r_{200}<0.07$ and (3) the virial ratio is
sufficiently close to unity, $2\,T/|U|<1.35$. These quantities provide a
measure for the dynamical state of a halo, and considering these three
conditions together guarantees in practice that a halo is close to an
equilibrium configuration, excluding the ones with ongoing mergers, or with
strong asymmetric configurations due to massive substructures.

For all relaxed halos selected in this way, we computed a spherically averaged
density profile by storing the halo mass in equally spaced bins in $\log_{10}
(r)$ between the virial radius $r_{200}$ and $\log_{10} (r/r_{200}) =
-2.5$. We used 32 bins for each halo and we choose a uniform radial range in
units of $r_{200}$ for the fitting procedure so that all halos are treated
equally, regardless of the mass. We find that we obtain stable results when we
use halos with more than 3000 particles, consistent with the \cite{Power2003}
criteria, while with fewer particles we notice resolution effects in the
concentration measurements, as both the gravitational softening and
discreteness effects can artificially reduce the concentration. The final mass
range we explored is hence $10^{12}$ to $10^{15}\,h^{-1}{\rm M}_{\odot}$.

\begin{figure}
\centering
\includegraphics[width=0.46\textwidth]{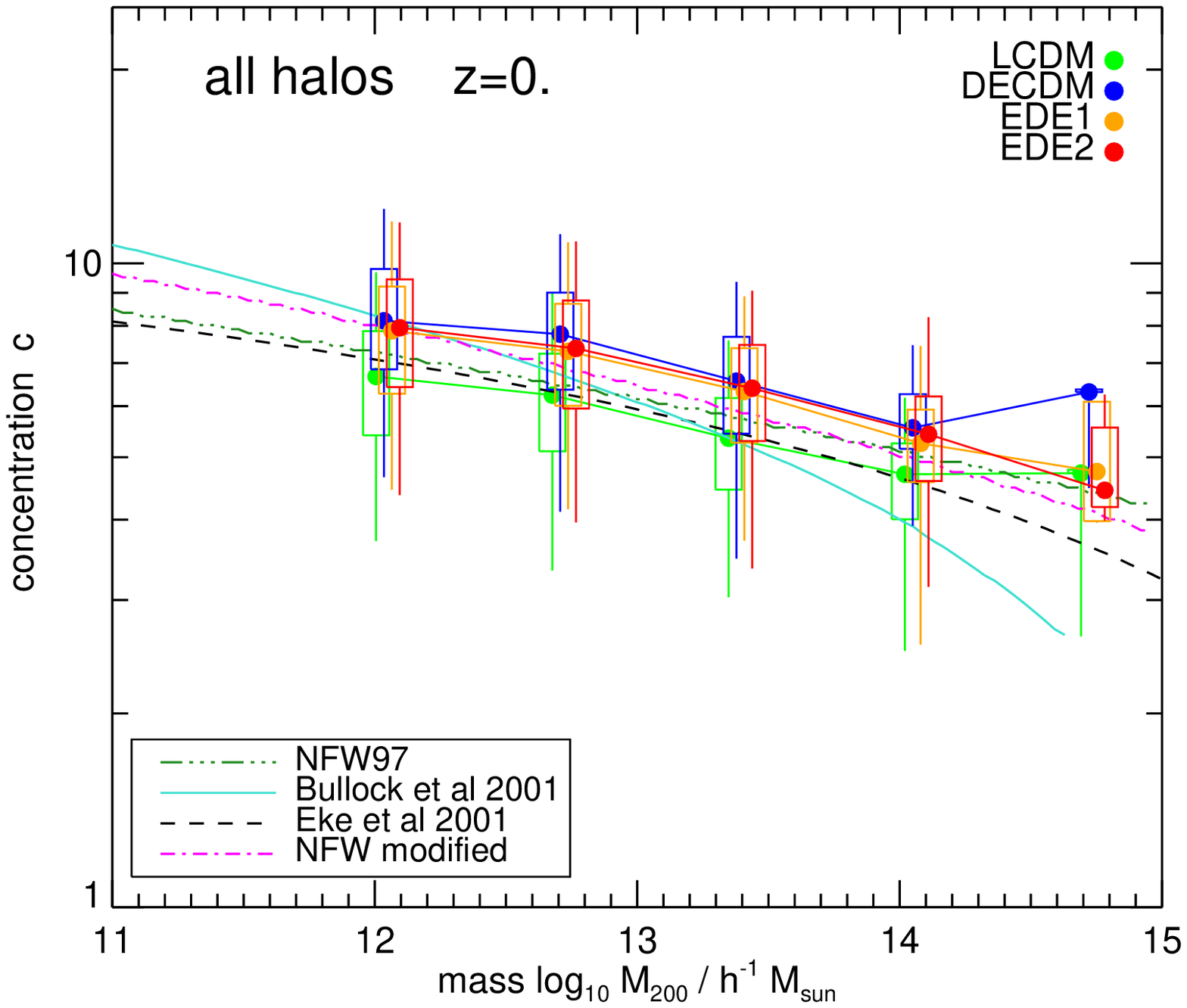}\\
\includegraphics[width=0.46\textwidth]{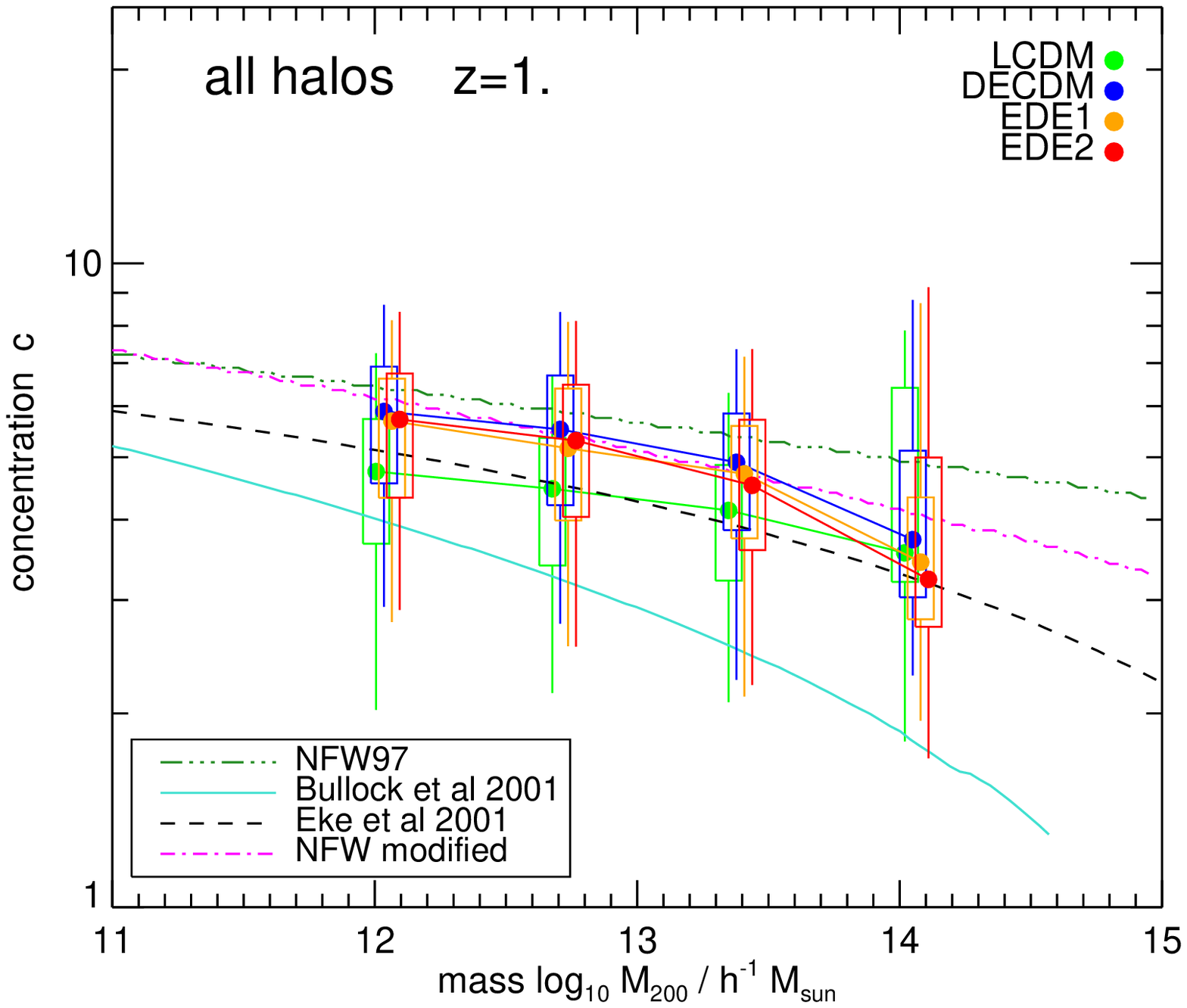}\\
\caption{Mass-concentration relation for all halos in our simulations. The
  top panels refers to redshift $z=0$, while the bottom panel shows the results at
  $z=1$. The boxes represent the 25 and 75 percentiles of the distribution with
  respect to the median value, while the whiskers show the 5 and 95
  percentiles. We compare our results with the theoretical expectations from
  NFW, ENS, B01. Also, a modified NFW prescription with slightly modified
  parameters as updated by \citet{Gao2008} is shown (see Section~5 for
  details). The concentration is $~5\%$ lower with respect to the relaxed
  sample at $z=0$ for the $\Lambda$CDM model.}
\label{fig:concentrationall}
\end{figure}

In Figure \ref{fig:concentration}, we show our measured mass-concentration
relation for the different dark energy models at $z=0$.  The four solid lines
show the mean concentration as a function of mass. The boxes represent the 25
and 75 percentiles of the distribution, while the whiskers indicate the 5 and
95 percentiles of the distributions. We note that the scatter of the
concentration at a given mass is very close to a log-normal distribution.  It
is interesting to remark that both the mean and the dispersion decrease with
mass. In fact, massive halos form in some sense a more homogeneous population,
because they have collapsed recently and so the formation redshift is
relatively close to the present epoch.  On the other hand, less massive halos
have a wider distribution of assembly redshifts and the structure of
individual objects strongly depends on their particular accretion histories.
For them, the assumption that objects we observe are just virialized is
therefore inappropriate, especially for very low mass halos. In
Fig.~\ref{fig:concentration} we take into account only the relaxed halos, but
we did an analogue measurement also for the whole sample, shown in
Figure~\ref{fig:concentrationall} at redshift $0$ (top panel) and at $z=1$
(low panel).

The correlation between mass and concentration approximately follows a power
law for the relaxed halos of the $\Lambda$CDM model.  In the literature, the
concentrations would be expected to be somewhat lower if a complete sample is
considered that includes disturbed halos. Comparing Figures
\ref{fig:concentration} and \ref{fig:concentrationall} we notice that this
expectation is confirmed, but the difference is not very pronounced, only about
$5\%$ for the whole mass range. We also note that the normalization
$\sigma_{8}=0.8$ used for our simulations slightly lowers the amplitude of the
zero point of the relation \citep{Maccio2008} when compared to the WMAP-3
normalization, as halos tend to assemble later with lower $\sigma_8$ and/or
$\Omega_m$.

When we compare our four simulated cosmologies we find that, as expected, EDE
halos of given mass have always higher concentration at a given redshift than
models with a cosmological constant: they tend to form earlier and so they
have a higher characteristic density.  Nevertheless, the differences are not
large, they deviate by no more than $\sim 27\%$ at $z=0$ over the entire mass
range we studied for all halos and $\sim 25\%$ for the relaxed one.  At higher
redshift, the differences are only slightly bigger, of order of $\sim 28\%$ at
$z=1$ for the whole sample, and $\sim 35\%$ for halos in equilibrium
configuration, suggesting that we anyway need reliable numerical calibrations
and highly accurate observational data to discriminate between the different
cosmologies.  Interestingly, the average concentration is almost independent
of mass when we consider $z \geq 2$, as the average concentration of the more
massive halos is similar at all redshifts \citep{Gao2008} and we are then
restricted to the exponential tail of the mass function.

The change in concentration normalization relative to the cosmological
constant model is well represented by the ratio between the linear growth
factor of different models at very high redshift,
\begin{equation}
c_{0} \rightarrow c_{0, \Lambda{\rm CDM}}\frac{D_{+}(\infty)}{D_{+,\Lambda {\rm CDM}}(\infty)},
\end{equation}
as suggested by \cite{Dolag2004}. In Table~\ref{tab:concentration}, we compare
the ratio between the concentration at $z=0$ both for the relaxed halos
(second column) and for the whole sample (third column) with the ratio between
the asymptotic growth factor for the same cosmologies (forth column). The
order of magnitude of the two effects is comparable, although the match is not
perfect. Here the ratios are computed for $M \sim 4 \times 10^{12} ~h^{-1}
M_{\odot}$, where we have a large number density of halos.

\begin{table}
\begin{center}

\begin{tabular}{l|rrr}
\hline
  Model       & $ \frac{c_{0}}{c_{0,\Lambda{\rm CDM}}}$ &
  $ \frac{c_{0,ALL}}{c_{0,ALL,\Lambda{\rm CDM}}}$ &
  $\frac{D_{+}(\inf)}{D_{+,\Lambda{\rm CDM}}(\inf)}$ \\

\hline
$\Lambda$CDM & 1.000 & 1.000 & 1.000    \\
DECDM        & 1.256 & 1.275 & 1.228    \\
EDE1         & 1.218 & 1.232 & 1.229    \\
EDE2         & 1.255 & 1.273 & 1.252   \\
\hline
\end{tabular}
\end{center}
\caption{Concentration and asymptotic growth factor in the four
  different cosmologies studied here. The ratio between the
  concentration parameters at redshift $z=0$ refers to a mass of $M \sim
  4 \times 10^{12} ~h^{-1} M_{\odot}$ that corresponds to the mass range that
  contains the majority of our halos. For each model (first column) we
  give the $c_0$ parameter relative to $\Lambda$CDM, taking into account
  the relaxed halos (second column) or the whole sample (third
  column). Finally, in the last column we show the linear growth factor
  at infinity relative to the $\Lambda$CDM cosmology.}
\label{tab:concentration}
\end{table}

It is interesting to compare the concentrations we measure with the various
theoretical predictions that have been made for this quantity. We investigate
three popular descriptions for the concentration: the classic Navarro, Frenk
\& White model (hereafter NFW), the model of \citet[][hereafter
B01]{Bullock2001}, and that of \citet[][hereafter ENS]{ENS2001}. Finally, we
also plot the new modified version of the original Navarro Frenk and White
model, as recently proposed by \citet{Gao2008}.  Both the B01 model and the
standard NFW have two free parameters that have been tuned to reproduce
simulation results.  In the original NFW prescription, the definition of the
formation time of a halo is taken to be the redshift at which half of its mass
is first contained in a single progenitor: $F=0.5$. The second parameter is
the proportionality constant, $C=3000$, that relates the halo density scale to
the mean cosmic density at the collapse redshift $z_{\rm coll}$.  Recently,
\citet{Gao2008} noticed that the evolution of the mass-concentration relation
with redshift can be approximated much better by setting $F=0.1$.  The B01
model adopts as collapse redshift the epoch at which the typical collapsing
mass fulfills $M_{*}(a_c)=F\,M_{\rm vir}$, with $F=0.01$. They further assume
that the concentration is a factor $K=3.4$ times the ratio between the scale
factor at the time the halo is identified and the collapse time. For $K$ and
$F$ we use the values that are indicated as the best parameters by
\citet{Maccio2007}.  Finally, we compute the ENS prescriptions considering the
effective amplitude of the power spectrum at the scale of the cluster
mass. This quantity, rescaled for the linear growth factor of the simulated
cosmology, has to be constant.  In this case, only one parameter, $C_{\sigma}=28$,
is needed.  \citet{Bullock2001} and \citet{ENS2001} refer to the virial radius
as the one including an overdensity given by the generalized top hat collapse
model. We have appropriately adapted these models such that the concentration
of a halo is defined instead relative to radius $r_{200}$, as in the NFW
model.

Aside from B01, all three other model predictions yield concentrations that
agree reasonably well with the measured values at $z=0$. The B01 model
underpredicts the relation at high masses, where it gives has a sharp decline
of the relation for $M \> 10^{13}\,h^{-1}{\rm M}_{\odot}$ which is not seen in
the simulations. In contrast, the NFW model is in reasonable agreement with
the data at $z=0$ for both halo samples. However, at $z=1$ the evolution
predicted by the NFW model is less than what we find numerically, even when we
consider the revised formulation proposed by \citet{Gao2008} (indicated as NFW
modified). The NFW model with the new fitting parameters yields a reasonable
fit at the high mass end, but performs a bit worse than the original
formulation at $z=0$, specially at low masses.  Unfortunately, for the NFW
model the normalization is model dependent, so we cannot really capture all
the effects due to different cosmological parameters we use.  Finally, the
dashed black line in each plot shows the ENS model. This prescription gives
the best match with our results and has been able to reproduce the slope of
the concentration-mass relation even at higher redshift.

\begin{figure}
\begin{center}
\includegraphics[width = 0.45\textwidth]{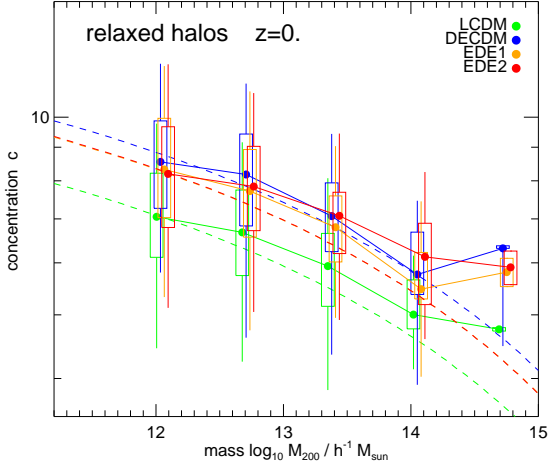}
\caption{Mass-concentration relation for relaxed halos today. Here we show the
  agreement between simulation results (symbols) and theoretical predictions
  from ENS (dashed lines), both for the $\Lambda$CDM and EDE cosmologies. To
  this end we solve Eqn.~(13) and (16) of Eke et al. (2001). The differences
  between the four cosmologies are due mostly to the differences in the growth
  factor evolution and consequently in the amplitude of the power
  spectrum. The ENS formula works quite well also for EDE models without
  modifications of the original prescription.}
\label{fig:concentrationENS}
\end{center}
\end{figure}

At a fixed mass, halos in the EDE cosmology are significantly less
concentrated than their counterparts in the $\Lambda$CDM cosmology.  It is
interesting to notice that the ENS model reproduces these differences quite
well, without modifications of the original prescription. In
Figure~\ref{fig:concentrationENS}, we plot for each simulation the
corresponding theoretical expectation (dashed lines) for the sample of relaxed
halos at $z=0$. For a low density universe the scaling of the linear growth
factor with redshift leads to a greater difference between the models. Dark
halo concentrations depend both on the redshift evolution of $\delta_c$ and
the amplitude of the power spectrum on mass scales characteristic for the
halo.

These results for the concentration are particularly important since they
demonstrate that quintessence cosmologies with the same equation-of-state at
present, but different redshift evolution, can produce measurable differences
in the properties of the non-linear central regions of cluster-sized
halos. However, the prospects to observationally exploit these concentration
differences to distinguish different dark energy cosmologies are sobering.
For one, the systematic differences we measure for the concentrations are
quite small compared to the statistical errors for the mean concentration,
while at the same time the theoretical algorithms for predicting the halo
concentration perform quite differently already for the $\Lambda$CDM
cosmology. Furthermore, directly measuring halo concentrations in observations
is not readily possible as it requires an accurate knowledge of the virial
radius of a halo, a parameter which is poorly constrained from
observations. It therefore remains to be seen whether the effects of dark
energy on the non-linear structure of dark halos can be turned into a powerful
tool to learn about the nature of dark energy.

\section{Counting halos by velocity dispersion}

As we have seen, the different evolution of the halo mass function is in
principle a very sensitive probe of the expansion history of the universe,
especially when the massive end of the mass function is probed. Obtaining
absolute mass estimates from observations is however problematic, and fraught
with systematic biases and uncertainties. It is therefore important to look
for new ways to count halos which are more readily accessible by observations.
 
One such approach lies in using the motion of galaxies in groups or clusters
of galaxies to measure the line-of sight velocity dispersion, which in turn
can be cast into an estimate of the total virial mass of the host halo. This
relies on the assumption that the dynamics of the cluster or group galaxies is
tracing out the dark matter halo potential.

Cluster and group galaxies can be identified with dark matter sub-structures
in N-body simulations \citep{Springel2001,Vale2004}. Employing the bulk
velocities of sub-halos as a simulation proxy for real galaxy velocities, we
can hence build a velocity profile for any isolated halo, and estimate a
line-of-sight velocity dispersion, similarly as it is done for observed group
catalogues of galaxies.  This allows then to directly count halos (i.e. galaxy
groups) as a function of line-of-sight velocity dispersion, bypassing the
problematic point of assigning halo mass estimates.

\begin{figure}
\begin{center}
\includegraphics[width = 0.4\textwidth]{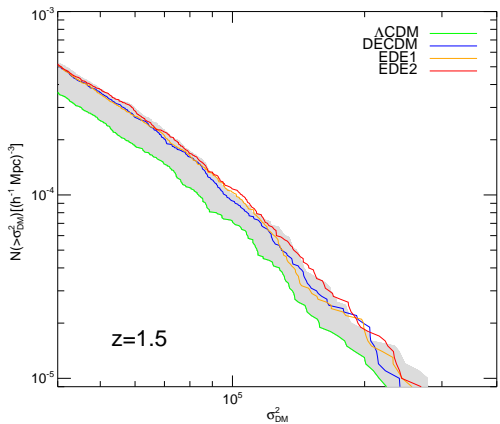}
\caption{The velocity function $n(\sigma)$ as a function of halo mass, for all
  satellites inside $r_{200}$. The shaded area indicates the differences
  between a $\Lambda$CDM model with $ \sigma_8=0.8$ and the same model with
  $\sigma_8=0.9$. It is interesting to remark that this EDE models could
  justify a higher normalization cosmology.}
\label{fig:Cumsigma2}
\end{center}
\end{figure}

In Figure~\ref{fig:Cumsigma2}, we show our estimated cumulative velocity
dispersion function for our four different cosmologies at redshift
$z=1.5$. This graph can be interpreted as being a different representation of
the halo mass function, except that it is in principle directly accessible by
observations.  For this measurement, we have derived the information on the
velocities from the {\small SUBFIND} algorithm directly implemented in {\small
  GADGET-3}, which can find subhalos embedded in dark matter halos.

An important aspect of this statistic is that it does not rely on the often
ambiguous definition of a group mass. Instead, it can be directly measured and
is more readily accessed by observations. In fact, studies based on the DEEP2
survey \citep{Davis2004,Davis2005,Conroy2007} indicate that, if combined with
both the velocity dispersion distribution of clusters from the Sloan Digital
Sky Survey and independent measurements of $\sigma_8$, they will be able to
constrain $w$ to within approximately $1\%$ accuracy.  This method is almost
independent of cosmological parameters, with the exception of $\sigma_{8}$,
since a change in normalization can shift the space density of halos as a
function of mass by a similar amount as done by the EDE models.  This is
illustrated in Figure~\ref{fig:Cumsigma2} by the shaded area, which represents
the change of the velocity dispersion function when $\sigma_8$ is increased
from $0.8$ (green line) to $0.9$ (upper limit of the shaded area).  The
velocity distribution function of the EDE models then approaches the one that we
would measure for a $\Lambda$CDM model with higher $\sigma_8$.

\begin{figure*}
\centering
\includegraphics[width=0.45\textwidth]{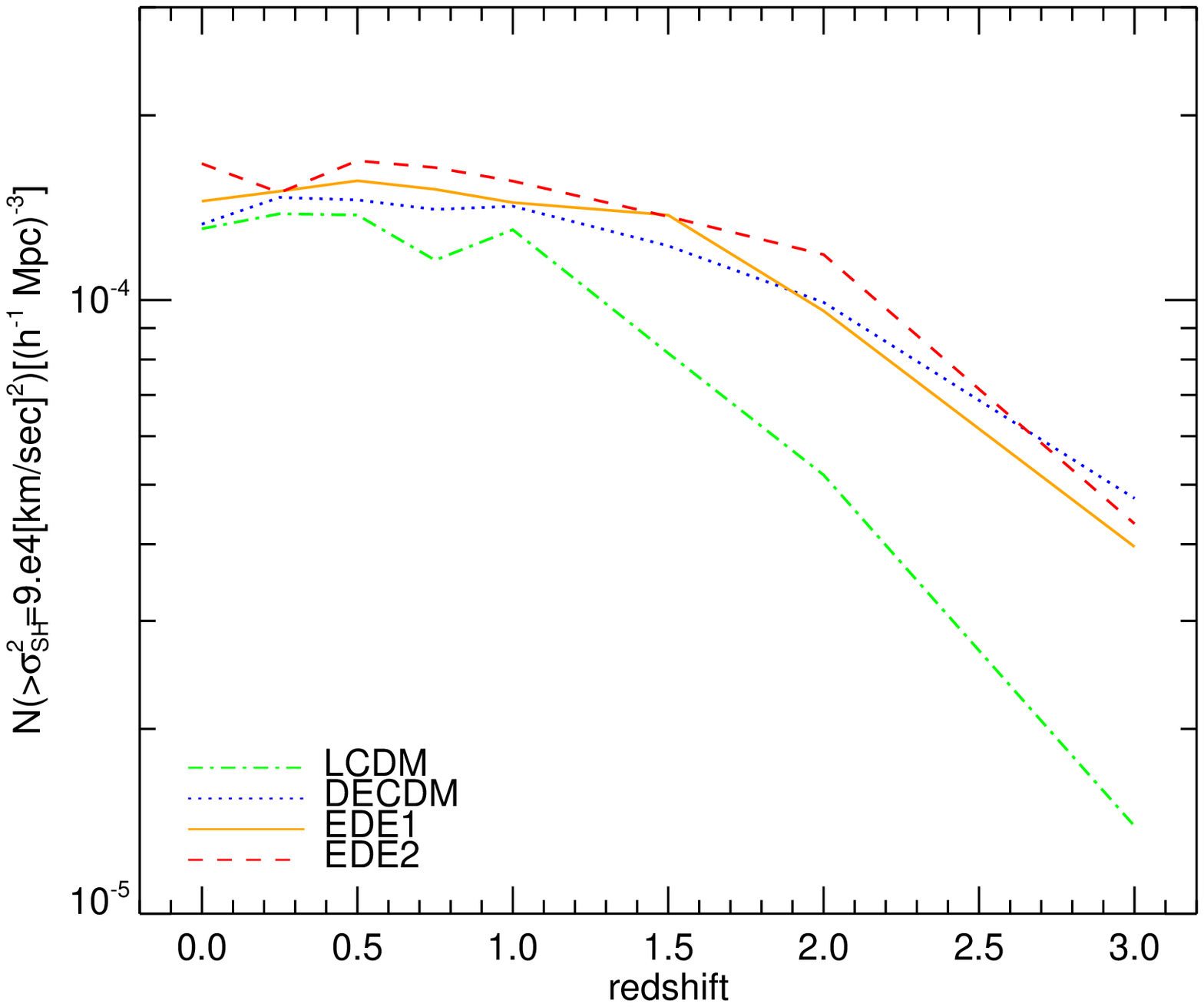}
\hfill
\includegraphics[width=0.45\textwidth]{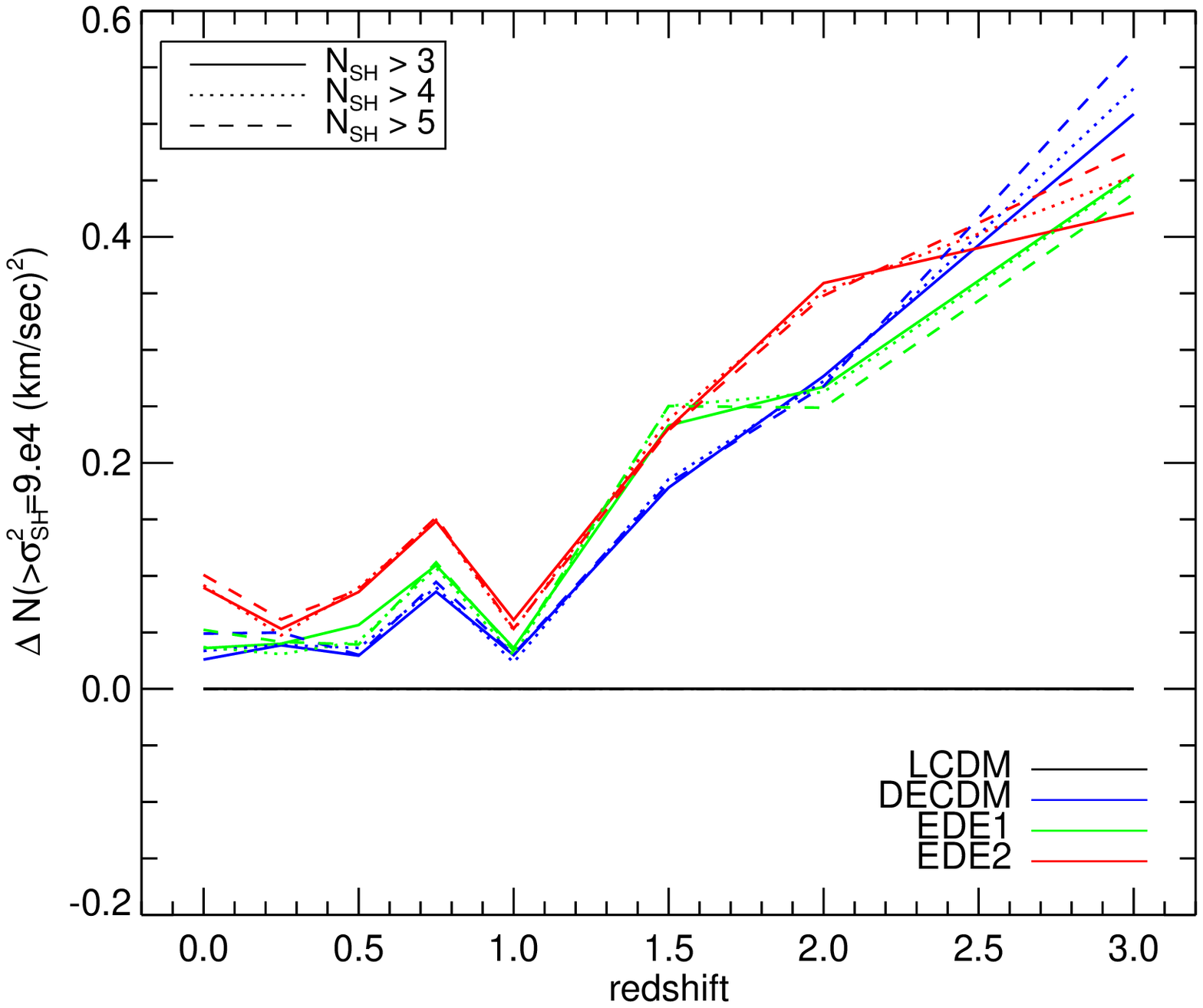}
\caption{Left panel: Comparison of the redshift evolution of the velocity
  dispersion function for all four cosmologies we simulated ($\Lambda$CDM,
  DECDM, EDE1, and EDE2). Here the cumulative count of groups with velocity
  dispersion above $\sigma = 300\,{\rm km\, s^{-1}}$ was used to measure the
  amplitude of the velocity dispersion function. Right panel: Differences in
  the number count when only halos with more than 3 (solid line), 4 (dotted
  line) or 5 (dashed line) substructures are selected. }
\label{fig:Cumsigma29.e4}
\end{figure*}

These kind of studies have strong motivations both from the observational and
theoretical point of view: there is little scatter between host galaxy
luminosity and dark matter halo virial mass and the velocity difference
distribution of satellites and interlopers can be modeled as a Gaussian and a
constant, respectively \citep{Conroy2005,Faltenbacher2006}.

Figure \ref{fig:Cumsigma29.e4} (left panel) shows the cumulative number of
groups with velocity dispersion above a given value, as a function in redshift
for the different models. We decided to count halos above a velocity
dispersion of $300\,{\rm km\,s^{-1}}$, where accurate measurements can be
expected also from observations.  Note that there is already a very large
difference between $\Lambda$CDM and EDE at redshift $z=1$.  We find that there
is almost no evolution in the cluster number in the dark energy models, while
$\Lambda$CDM drops by a factor of nearly 10 up to redshift $z=3$.  What is
especially important here is the relative difference between the number counts
of the two different cosmologies. The fact that we do not need to introduce
the mass in this comparison give us the advantage of having no error derived
from the particular measurement procedure adopted for the mass.  At a fixed
velocity dispersion, we can directly probe the growth of the structure at each
redshift, which depends on the equation of state parameter $w$. The slower
evolution of the cluster population in EDE models is exactly what is expected
to be observed also from Sunyaev-Zeldovich studies of large samples of
clusters of galaxies.  Combined with probes of the cluster internal velocity
dispersion we can hence hope to be able to derive stringent cosmological
constraints.
 
We also remark that the relative difference between the number of objects
within these four simulations seems to be a quite robust statistic which is
invariant with respect to details of the measurement procedure. For example,
in Figure~\ref{fig:Cumsigma29.e4} (right panel), we change the number of
considered subhalos in the halos to be a minimum of 3, 4, and 5, but the
velocity dispersion function relative to the $\Lambda$CDM cosmology is
essentially unchanged. In practice, the number of observable satellites per
host halo suffers from limitations imposed by the magnitude limit of the
survey. Our results suggest that the measured velocity dispersion should
however be relatively insensitive to this selection effect.

\begin{figure}
\begin{center}
\includegraphics[width = 0.4\textwidth]{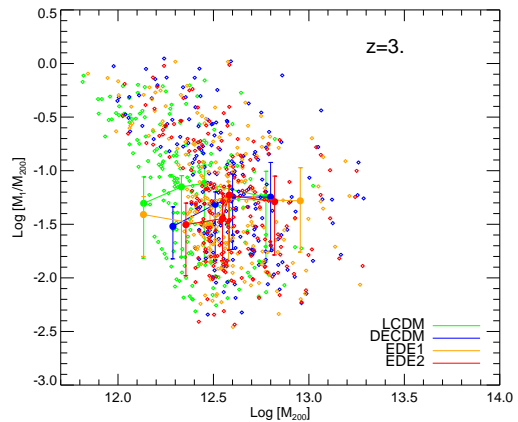}
\caption{Ratio of the mass of the two most massive substructures with respect
  to the mass of the parent halo. The small diamonds refer to individual
  halos, while the filled circles are the median values. The error bars mark
  the 20th and 80th percentiles of the distributions.}
\label{fig:subhalo}
\end{center}
\end{figure}

Finally, we have also studied a few properties of the largest substructures in
halos to see whether there is a difference in EDE cosmologies.  In
Figure~\ref{fig:subhalo}, the small diamonds indicate the values of the ratio
between $M_1$ (the mass of the most massive subhalo) and $M_{200}$ (the mass
within a sphere of density 200 times the critical value at redshift 0) for the
first 200 most massive halos at redshift $z=3$.  The filled circles represent
the median of the distribution, computed in bins of 50 halos each, while the
error bars mark the 20-th and 80-th percentiles of the distribution. There is
almost no dependence on parent halo mass, but we can notice a small, but
systematic tendency for the $\Lambda$CDM subhalos to be slightly more
massive. The dependence is quite weak, yet this behaviour is clear even if the
mass of the progenitor $M_{200}$ tends to be lower on average at this high
redshift.  This is symptomatic of the fact that the $\Lambda$CDM substructures
are formed at lower redshift with respect to what happens in the EDE
models. This is also consistent with expectations based on the observed
dependence of substructure mass fraction on halo mass
\citep[e.g.][]{DeLucia2004}.  Once accreted onto a massive halo, subhalos
suffer significant stripping, an effect that is more important for
substructures accreted at higher redshift, making the subhalos in the EDE
models less massive on average.

\section{Conclusions}

In this study we have analyzed non-linear structure formation in a particular
class of dark energy cosmologies, so called {\em early dark energy} models
where the contribution of dark energy to the total energy density of the
universe does not vanish even at high redshift, unlike in models with a
cosmological constant and many other simple quintessence scenarios. Our
particular interest has been to test whether analytic predictions for the halo
mass function still reliably work in such cosmologies. As the evolution of the
mass function is one of the most sensitive probes available for dark energy,
this is of crucial importance for the interpretation of future large galaxy
cluster surveys at high redshift. The mass function of EDE models is also
especially interesting because analytic theory based on extensions of the
spherical collapse model predicts that the mass function should be
significantly modified \citep{Bartelmann2006}, and in particular be
characterized by a different value of the linear overdensity $\delta_c$ for
collapse, as well as a slightly modified virial overdensity.

We have carried out a set of high-resolution N-body simulations of two EDE
models, and compared them with a standard $\Lambda$CDM cosmology, and a model
with a constant equation of state equal to $w=-0.6$.  Interestingly, we find
that the universality of the standard Sheth \& Tormen formalism for estimating
the halo mass function also extends to the EDE models, at least at the $\simlt
15 \%$ accuracy level that is reached also for the ordinary $\Lambda$CDM
model. This means that we have found good agreement of the standard ST
estimate of the abundance of DM halos with our numerical results for the EDE
cosmologies, {\em without} modification of the assumed virial overdensity and
the linear density contrast threshold.  This disagrees with the theoretical
suggestions based on the generalized top-hat collapse. In fact, if we instead
use the latter as theoretical prediction of the halo mass function, the
deviations between the prediction and the numerical results become
significantly larger.  We hence conclude that the constant standard value for
the linearly extrapolated density contrast can be used also for an analysis of
early dark energy cosmologies.  Very recently, similar results were also
obtained by \cite{Francis2008}, who studied the same problem in cosmological
simulations with somewhat smaller mass resolution.

This results on the mass function appear to hold over the whole redshift range
we studied, from $z=0$ to $z=3$. Since our simulations were normalized to the
same $\sigma_8$ today, their mass functions and power spectra agree very well
today, but towards higher redshift there are significant differences, as
expected due to the different histories of the linear growth factor in the
different cosmologies. In general, structure in the EDE cosmologies has to
form significantly earlier than in $\Lambda$CDM to arrive at the same
abundance today. For example, already by redshift $z=3$, the abundance of
galaxy clusters of mass $M=5\times 10^{12}\,h^{-1}{\rm M}_\odot$ is higher in
EDE1 by a factor of $\sim 1.7$ relative to $\Lambda$CDM.

The earlier formation of halos in EDE models is also directly reflected in the
concentration of halos. While for a given $\sigma_8$ we find the same
abundance of DM halos, the different formation histories are still reflected
in a subtle modification of the internal structure of halos, making EDE
concentrations for all halo masses and redshifts considered slightly
higher. The difference is however quite small, but it would, for example, lead
to a higher rate of dark matter annihilation in halos.

Another relationship that appears to accurately hold equally well in
$\Lambda$CDM as in generalized dark energy cosmologies is the virial scaling
between mass and dark matter velocity dispersion that \citet{Evrard2008} has
found. In fact, we find that their normalization of this relation is
accurately reproduced by all of our simulations within the measurement
uncertainties, independent of cosmology. This also suggests that possible
differences in the virial overdensity of EDE halos must be very small, and
that presumably the relationship between total Sunyaev-Zeldovich decrement and
halo mass is unmodified as well.
 
We show that counting the number of halos as a function of the line-of-sight
velocity dispersion (of subhalos or galaxies), both in simulations and
observations, can probe the growth of structures with redshift, and so put
powerful constraints on the equation of state parameters.  This goal can be
achieved by just identifying and counting groups in galaxy survey data such as
DEEP2, and by comparing them with high-resolution N-body
simulations. Precision measurements with this technique will still require
accurate calibrations to deal with complications such as a possible velocity
bias or selection effects in observational surveys.  However,
\citet{Davis2005} suggest that the DEEP2 survey alone has the power to
constrain $w$ to an accuracy of 20\% using velocity dispersion data, which
illustrates the promise of this technique. In combination with other
independent data, such as X-ray temperature and SZ decrement data, the
constraints could be improved to an accuracy of 5\%, without the need to
invoke a model for the ambiguous total mass of a halo.

Distinguishing a time-varying dark energy component from the cosmological
constant is a major quest of the present theoretical and observational
astronomy. One approach is to rely on classical cosmological tests of the
Hubble diagram, e.g.~by pushing the supernova type Ia observations to much
higher redshift. Another quite direct geometrical probe is the observation of
baryonic acoustic oscillations in the matter distribution at different
redshifts.  Finally, the linear and non-linear evolution of cosmic structures
provides another opportunity to constrain dark energy. In this work we have
used numerical N-body simulations to examine the difference in structure growth
in early dark energy cosmologies. We have seen that such simulations are
essential to test the predictions of more simplified analytic models, and to
calibrate observational tests that try to constrain the properties of dark
energy with the abundance and internal structure of dark matter halos. Our
results show clearly that the effects due to dynamical dark energy tend to be
quite subtle, and can only be cleanly distinguished from ordinary $\Lambda$CDM
in high accuracy simulations. This poses new challenges to improve the
precision of future generations of simulations, and at the same time
emphasizes the immense observational task to arrive at sufficiently precise
data at high redshift to constrain the dark side of the universe with the
required accuracy.

\section*{acknowledgements}

We acknowledge Matthias Bartelmann for providing us with his code LIBASTRO for
the virial overdensity prediction in EDE models. We also thank Klaus
Dolag for useful discussions. Computations were
performed on the OPA machine at the computer centre of the Max Planck Society.
This research was supported by the DFG cluster of excellence ‘Origin and
Structure of the Universe’.

\bibliographystyle{mn2e}
\bibliography{master}

\label{lastpage}

\end{document}